\documentclass[aps,pra,superscriptaddress,twocolumn]{revtex4-2}

\usepackage[normalem]{ulem}
\usepackage[american]{babel}
\usepackage{graphicx}
\usepackage{dcolumn}
\usepackage{bm}
\usepackage{hyphenat}
\usepackage{amssymb}
\usepackage{verbatim}
\usepackage{xpatch}
\usepackage{color}
\usepackage{epstopdf}
\usepackage[table,x11names]{xcolor}
\usepackage{float}
\usepackage[caption=false]{subfig}
\usepackage{soul}
\usepackage{multirow}
\usepackage{hyperref}
\usepackage{makecell}
\usepackage{cancel}
\usepackage{amsmath}
\usepackage{simpler-wick}
\hypersetup{
	colorlinks,
    linkcolor={blue!75!black},
	citecolor={blue!75!black},
	urlcolor={blue!75!black}
}

\graphicspath{{./figures/}}

\begin{document}

\title{Static dipolar impurity in a Fermi gas}

\author{Eugen Dizer}
\affiliation{Institut für Theoretische Physik, Universität Heidelberg, 69120 Heidelberg, Germany}

\author{Xin Chen}
\affiliation{Institut für Theoretische Physik, Universität Heidelberg, 69120 Heidelberg, Germany}

\author{Richard Schmidt}
\affiliation{Institut für Theoretische Physik, Universität Heidelberg, 69120 Heidelberg, Germany}

\author{Arthur Christianen}
\affiliation{Institute for Theoretical Physics, ETH Zürich, 8093 Zürich, Switzerland}


\begin{abstract}

We consider a static dipolar impurity immersed in a Fermi gas and show how dipolar interactions shape both the few-body scattering properties and the resulting polaron physics. Using Fumi's theorem, we connect the anisotropic phase shifts directly to the ground-state energy, and compute the exact impurity spectrum via the functional determinant approach. The resulting spectrum is a convolution of independent even- and odd-wave contributions, giving rise to a ground state with simultaneously occupied even- and odd-wave bound states whose separation is tunable via the dipole length. A variational ansatz limited to one particle-hole excitation fails to capture this doubly occupied ground state. We further find that long-range dipolar interactions weaken Anderson's orthogonality catastrophe, reducing the power-law decay of the quasiparticle weight. These effects can be tested with ultracold polar molecules, where microwave shielding and independent tunability of interactions enable experimental access to the predicted regimes.

\end{abstract}

\maketitle

\section{Introduction} \label{sec:Introduction}

Impurities in a quantum medium serve as a sensitive probe of few- and many-body physics~\cite{Schmidt2018,Massignan:2025uee}. The medium dresses the impurity with its excitations, forming a polaron whose properties are predominantly determined by the underlying two-body scattering. In strongly correlated Fermi systems, polarons provide a controlled quasiparticle picture of complex many-body physics~\cite{Massignan:2014,Scazza:2022bez}. A prominent example is the density-imbalanced unitary Fermi gas, where the normal phase is described as a Fermi liquid of polarons~\cite{Schirotzek:2009exp,Punk:2009,ChevyMora2010}.

Most studies so far have focused on short-range interactions between impurity and bath. Recent advances with magnetic atoms~\cite{Chomaz:2023} and, in particular, ultracold polar molecules~\cite{demarco:2019,schindewolf:2022,Bigagli:2024,schindewolf:2026} now open a qualitatively different regime governed by dipolar interactions. Dipolar interactions are anisotropic and long-ranged, coupling different partial-wave channels and producing strong finite-range corrections~\cite{Bohn:2009,Bohn:2014,PhysRevLett.108.215301,Baranov:2012}. This anisotropic, momentum-dependent scattering underlies a broad landscape of many-body phases, from supersolids and self-bound droplets to unconventional dipolar superfluidity~\cite{Lahaye:2009qqr,Baranov:2002,cooper:2009,levinsen:2011,deng:2023}.

Impurities with long-range interactions exhibit especially rich properties. For example, the polaron spectra of ionic~\cite{PhysRevA.105.023309} or Rydberg atom impurities \cite{Camargo:2018,Schmidt:2018rydberg,Whalen:2019,Sous2020,Gievers2024,Durst:2024ryd,Durst:2026anisotropic} show multiple bound states between the impurity and bath particles. Dipolar polarons have begun to be explored in both Bose and Fermi gases~\cite{PhysRevA.85.020702,PhysRevA.103.033324,Volosniev:2023xfn,Nakano:2024the,Tiene:2024,PhysRevA.89.023612,Shukla2024,Shukla:2025}, but these works focus on ground-state properties, lower dimensions or treat the impurity-bath coupling within mean-field or contact approximations. The full dipolar scattering structure between impurity and fermionic bath -- the bound states, anisotropic phase shifts, and their imprint on the polaron spectrum -- has remained unexplored.

In this work, we consider a static dipolar impurity in a non-interacting Fermi gas. Because the impurity is static, the Hamiltonian remains quadratic in fermionic operators and can be solved numerically exactly, connecting the two-body scattering directly to the polaron properties and to Anderson's orthogonality catastrophe~\cite{Anderson:1967}. This connection is made explicit by the generalized Fumi's theorem~\cite{Fumi:1955}, which relates the impurity-induced energy shift $\Delta E$ to the scattering phase shift~\cite{Mahan2000},
\begin{equation} \label{eq:generalized-Fumi}
    \Delta E = -\frac{1}{\pi} \int_0^{E_F} dE \, \delta(E) \,,
\end{equation}
where $E_F$ is the Fermi energy and $\delta(E)$ collects the contributions of all scattering channels via the $T$-matrix~\cite{Xiang2022},
\begin{align} \label{eq:generalized-Fumi-shift}
    \delta(E) = \text{Im}\log\det\,T(E) \,.
\end{align}
The anisotropy of the dipole potential couples different partial waves and separates the multichannel structure into distinct symmetry sectors~\cite{PhysRevA.80.052702,PhysRevA.81.042708,PhysRevA.107.033315,PhysRevA.99.032709}. We show how this modifies the polaron physics, and contrast the aligned (dipolar) and anti-aligned (anti-dipolar) configurations accessible with microwave-shielded molecules \cite{Karman2025DMS}.

The paper is organized as follows. In Section~\ref{sec:two-body-problem}, we characterize the dipolar scattering, including the bound-state structure and the modified effective-range expansion. In Section~\ref{sec:many-body-problem}, we first use the generalized Fumi's theorem and numerical phase shifts to determine exact polaron energies. We then compute the exact spectrum and show that it factorizes into even- and odd-wave contributions, giving a ground state with simultaneously occupied even- and odd-wave bound states. We benchmark the widely used Chevy ansatz~\cite{PhysRevA.74.063628} against the exact solution, and show how long-range interactions weaken the orthogonality catastrophe. We conclude in Section~\ref{sec:Conclusion} with a discussion of the implications for future studies of dipolar many-body systems. Throughout this work, we use natural units, in which $\hbar = k_B = 1$.

\begin{figure}[t]
    \centering
    \includegraphics[width=\linewidth]{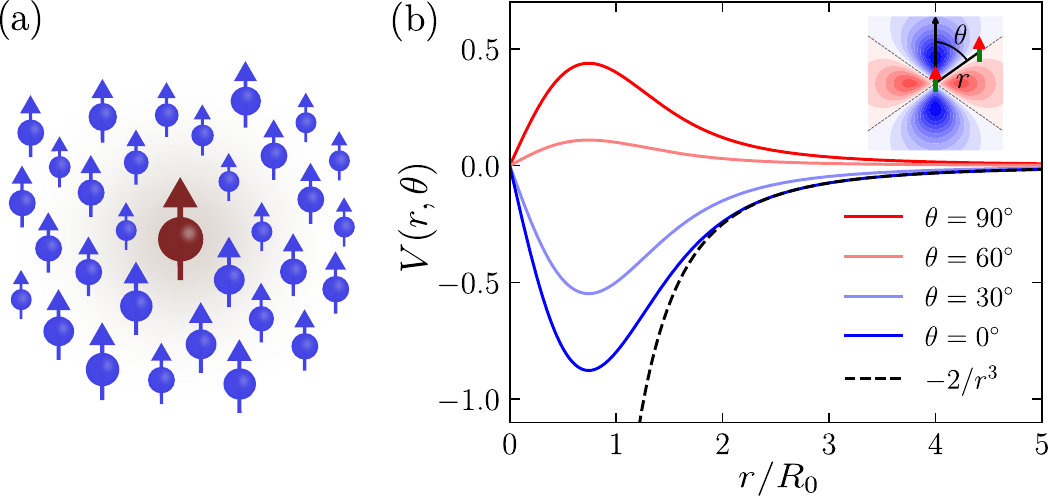}
    \caption{Static dipolar impurity in a Fermi gas. (a) A static dipolar impurity (red) is immersed in a three-dimensional Fermi gas (blue), and coupled via dipole-dipole interactions. (b) Regularized dipole--dipole potential $V(r,\theta)$ between the bath fermions and the impurity, shown for several angles $\theta$ to the polarization axis. The inset shows the model potential as a function of $r$ and $\theta$.}
    \label{fig:1}
\end{figure}
%

\section{Two-body properties} \label{sec:two-body-problem}

The many-body response of a Fermi gas to a dipolar impurity [Fig.~\ref{fig:1}(a)] is governed by the underlying two-body scattering properties. In this Section, we analyze the low-energy scattering problem associated with a regularized dipolar potential focusing on bound-state formation, scattering lengths, and the modified effective-range expansion. These results establish the microscopic input for the impurity spectra discussed in Sec.~\ref{sec:many-body-problem}. Although we focus on an infinitely heavy impurity for the polaron problem, the scattering discussion applies for arbitrary impurity mass.

We consider the relative motion of a fermion of mass $m$ scattering off an impurity of mass $M$ in three dimensions. In the center-of-mass frame, the two-body Schrödinger equation reads
\begin{align} \label{eq:Schrödinger-equation-regulated}
    \left[-\frac{\nabla^2}{2m_r} + V(\bm{r})\right]\psi(\bm{r}) = E\psi(\bm{r}) \,,
\end{align}
where $m_r=mM/(m+M)$ is the reduced mass.

To theoretically describe the dipolar interaction, we need to regularize the singular short-distance behavior of the bare dipole–dipole potential~\cite{PhysRevLett.110.045302}. We introduce a model potential,
\begin{align} \label{eq:model-potential-real}
    V(\bm{r}) = \frac{d^2}{r^3} \left(1-3\cos^2\theta\right) g(r) \,,
\end{align}
where $d$ is the dipole moment, $\theta$ denotes the angle between $\bm{r}$ and the polarization axis, and the function $g(r)=(1-e^{-r^2/R^2_0})^2$ introduces a short-range cutoff $R_0$. This regularization removes the divergence at short distances while preserving the asymptotic dipolar tail and the angular structure. The form of $g(r)$ is chosen such that $V(\bm{r})$ has a well-behaved Fourier transform with a compact analytic expression, given in Appendix~\ref{app:chevy-ansatz}. The regularized dipole potential is shown in Fig.~\ref{fig:1}(b). The specific choice of short-range regularization does not affect the universal low-energy scattering properties, which are controlled by the scattering length $a$ and the dipole length $a_d$ through the modified effective-range expansion discussed in Sec.~\ref{subsec:phase-shift}. Corrections appear only at higher order via finite-range effects. Alternative forms of regularized potentials have been employed in previous work~\cite{Roudnev:2009,Wang2012,PhysRevLett.110.045302}.

In addition to aligned dipoles, we also consider the case of anti-dipolar interactions, corresponding to imaginary values of $d$. This situation is of particular relevance for ultracold polar molecules, where microwave dressing and shielding schemes allow to realize both dipolar and anti-dipolar interactions~\cite{PhysRevLett.121.163401,Chen:2022svd,Chen:2023dqs,Karman2025DMS,Deng:2025fjv,schindewolf:2026}.

We introduce dipole units~\cite{Bohn:2014},
\begin{align} \label{eq:dipole-units}
    a_d = m_r d^2 \,, \quad E_d = \frac{1}{m_r^3 d^4} \,,
\end{align}
which define the natural length and energy scales of the problem. Expressed in dimensionless variables, $r\rightarrow r/a_d$ and $E \rightarrow E/E_d$, the Schrödinger equation becomes 
\begin{align} \label{eq:Schrödinger-equation-dimensionless}
    \left[ -\frac{1}{2}\nabla^2 -\frac{2 C_{20}}{r^3} g(r) \right]\psi(\bm{r}) = E \psi(\bm{r}) \,,
\end{align}
where $C_{20}=(3\cos^2\theta-1)/2$ is the standard reduced spherical harmonic~\cite{Bohn:2009}. To proceed, we expand the dimensionless wave function into partial waves,
\begin{align} \label{eq:partial-waves}
    \psi_m(r,\theta,\phi) = \frac{1}{r} \sum_{l} F_{lm}(r) Y_{lm}(\theta,\phi) \,.
\end{align}
Since the magnetic quantum number $m$ is conserved, each sector can be treated independently, leading to a set of coupled radial equations for fixed $m$. Additionally, the even and odd angular-momentum channels decouple because of the even parity of the dipole potential, allowing us to analyze the even- and odd-parity scattering and bound states separately (see Appendix~\ref{app:dipolar-scattering} for details). The radial wave functions $F_{lm}(r)$ satisfy the coupled-channel equations,
\begin{align} \label{eq:coupled-channel-Schrödinger-equations}
    -\frac{1}{2}\frac{d^2}{dr^2}F_{lm} + \frac{l(l+1)}{2r^2}F_{lm} - \frac{2g}{r^3} \sum_{l'} C^m_{ll'} F_{l'm} = E F_{lm} \,,
\end{align}
where the coupling matrix elements $C^m_{ll'}$ are given by~\cite{Bohn:2009}
\begin{align} \label{eq:coupling-matrix-element}
    C^m_{ll'} &= \langle lm|C_{20}|l'm\rangle \notag \\ &= (-1)^m \sqrt{(2l+1)(2l'+1)} \notag \\
    &\quad \times
    \begin{pmatrix}
        l & 2 & l' \\
        -m & 0 & m
    \end{pmatrix}
    \begin{pmatrix}
        l & 2 & l' \\
        0 & 0 & 0
    \end{pmatrix} \,,
\end{align}
using $3j$ symbols. Note that $C_{00}^0=0$, implying that there is no direct interaction in the $s$-wave channel in the Born approximation. However, even the pure dipole potential allows to define a finite $s$-wave scattering length through couplings with other partial waves at higher orders in the Born expansion via the $C_{02}^0$ coefficient~\cite{Bohn:2009}. Instead, for $p$-wave scattering there is a direct coupling already at first order, with $C_{11}^0=2/5$ and $C_{11}^{\pm1}=-1/5$.

\subsection{Low-energy scattering} \label{subsec:scattering}

Due to the long-range nature of dipolar interactions, the low-energy scattering properties are substantially different from the contact interactions often studied in quantum impurity problems. In the limit of zero collision energy, the scattering amplitude takes the form~\cite{Bohn:2014}
\begin{align}
    f(\mathbf{k},\mathbf{k}') = -a - \frac{2a_d}{3} \left(3\cos^2\theta_{\mathbf{\mathbf{q}}} - 1\right) \,,
\end{align}
where $\mathbf{k}$ and $\mathbf{k}'$ denote the initial and scattered wave vector, respectively, $a$ is the $s$-wave scattering length, and $\theta_{\mathbf{\mathbf{q}}}$ is the angle between the momentum transfer $\mathbf{q}=\mathbf{k}-\mathbf{k}'$ and the polarization axis. Although this expression is derived within the first-order Born approximation, it well represents the scattering properties up to renormalization of the parameters~\cite{oldziejweski:2016}. However, we work with the full numerical scattering amplitude in the following.

At finite collision energy $E=k^2/(2m_r)$, the scattering amplitude $f(\mathbf{k},\mathbf{k}')$ is given by the on-shell $T$-matrix elements $T^m_{ll'}(k)$, with $k=|\mathbf{k}|=|\mathbf{k}'|$ fixed on the energy shell~\cite{Bohn:2009},
\begin{align}
    f(\mathbf{k},\mathbf{k}') = -\frac{2\pi}{k} \sum_{ll'm} i^l
    Y^*_{lm}(\hat{\mathbf{k}})\, T^m_{ll'}(k)\, i^{-l'}
    Y_{l'm}(\hat{\mathbf{k}}') \,,
\end{align}
where the $T$-matrix is connected to the $S$-matrix via $T(k)=i[S(k)-I]$, with identity matrix $I$.

To relate the microscopic model potential of Eq.~\eqref{eq:model-potential-real} to the physical observables, we calculate the scattering length $a$ using numerical coupled-channel calculations. The $s$-wave scattering length $a$ is extracted from the $l=m=0$ component~\cite{PhysRevA.75.052705},
\begin{align}
    a = \lim_{k\rightarrow 0} \frac{T^0_{00}(k)}{2k}  \,.
\end{align}
We note that the ratio $T^0_{00}(k)/(2k)$ is complex in general. However, the imaginary part of the $T$-matrix vanishes asymptotically as $k\rightarrow 0$, such that the scattering length $a$ is purely real~\cite{Wang2012}.

\begin{figure*}[t]
    \centering
    \includegraphics[width=\linewidth]{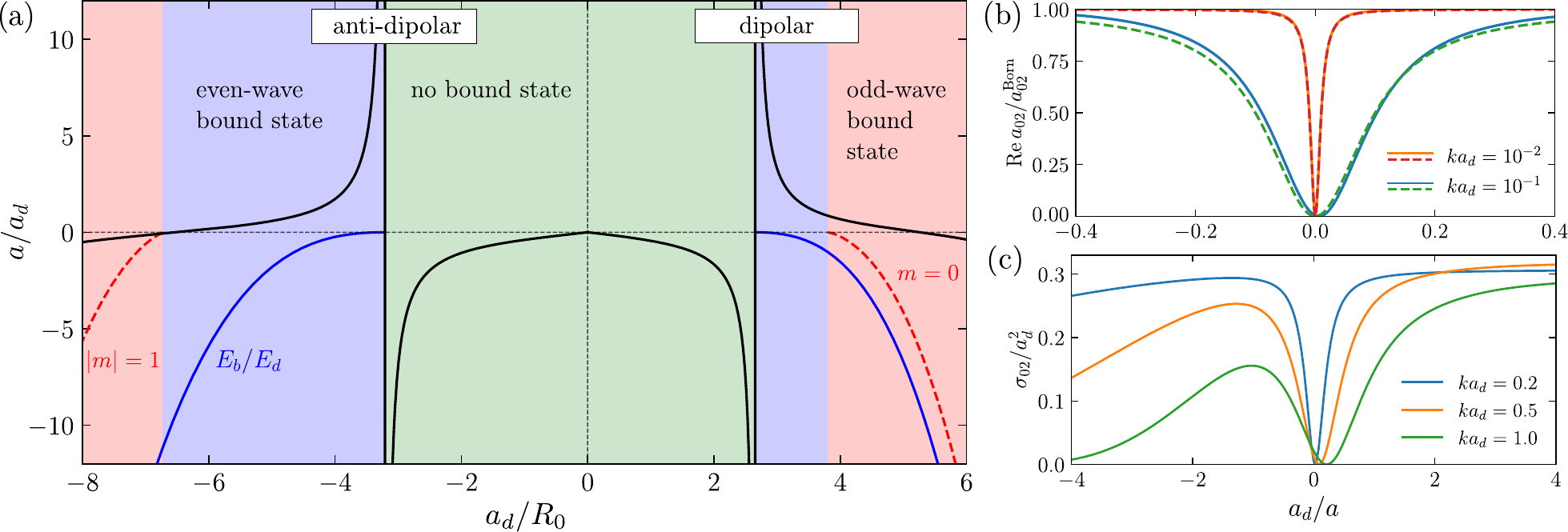}
    \caption{(a) The $s$-wave scattering length $a/a_d$ as a function of $a_d/R_0$. Regions without bound states (green), with an even‑wave bound state (blue), and with an odd‑wave bound state (red) are highlighted. The corresponding bound‑state energies $E_b/E_d$ are indicated by blue solid (even‑wave) and red dashed (odd‑wave) curves. For $a_d>0$, the odd‑wave bound state belongs to the $m=0$ angular‑momentum sector, whereas for $a_d<0$, it appears twice in the $|m|=1$ sector. (b) The real part of the dipole scattering length $\text{Re}\,a_{02}/a^{\mathrm{Born}}_{02}$ as a function of $a_d/a$ for different collision energies $ka_d$. A characteristic dip appears at the $s$-wave resonance. The behavior is well captured by the analytic expression in Eq.~\eqref{eq:dipolar-length} (dashed lines), which agrees closely with the numerical results (solid lines). (c) The cross section $\sigma_{02}/a^2_d$ as function of $a_d/a$ for larger collision energies $ka_d$ shows clear deviations from the universal expression in Eq.~\eqref{eq:cross-section}.}
    \label{fig:2}
\end{figure*}

The resulting scattering length $a/a_d$ is shown in Fig.~\ref{fig:2}(a) as a function of inverse short-range cutoff $a_d/R_0$. The figure is divided into three regions according to the presence of bound-states. For weak dipolar coupling, no bound state is present (green region). Upon increasing $|a_d|/R_0$, an even-wave bound state emerges first (blue region), followed by an odd-wave bound state at stronger coupling (red region), with the corresponding bound-state energies $E_b/E_d$ indicated by solid and dashed lines, respectively. It can be shown that the even-wave bound state always appears lower in energy than the odd-wave bound state, for both dipolar and anti-dipolar potentials. The specific value of $a/a_d$ at which the odd-wave bound state appears, however, is non-universal and depends on the short-range potential. For anti-aligned dipoles, corresponding to a sign-reversed dipolar interaction, the onset of bound states is shifted.

A crucial difference between dipolar and anti-dipolar scattering is the $m$-sector of the odd-wave bound state(s), see Fig.~\ref{fig:2}(a). For positive $a_d$, the odd-parity bound state lies in the $m=0$ angular-momentum channel, whereas for negative $a_d$ we find two degenerate bound states in the $|m|=1$ sector. Moreover, for anti-dipolar interactions, the odd-wave bound state emerges only at significantly larger values of $a_d/R_0$. This behavior can be understood from the diagonal matrix elements $C^m_{11}$ within the Born approximation. While $C^0_{11}=2/5$, the $|m|=1$ channel yields $C^1_{11}=-1/5$, implying that for anti-dipolar interactions the $|m|=1$ sector is attractive but with only half the effective strength. Consequently, a larger dipole moment is required to support a bound state in this channel.

Aside from the $s$-wave scattering length, we can also extract the anisotropic part of the scattering amplitude from the coupled-channel calculations. We define
\begin{align}\label{eq:a02}
    a_{02}(k) = \frac{T^0_{02}(k)}{2k} \,.
\end{align}
In Fig.~\ref{fig:2}(b), we show the real part of $a_{02}$ as a function of inverse scattering length $a_d/a$ for two values of the scattering momentum $k$. Similar to the case of the $s$-wave scattering length, the imaginary part of this $T$-matrix element vanishes as $k\rightarrow 0$, such that $a_{02}$ is real in this limit. We find that the value of $\text{Re}\,a_{02}$ at small $k$ agrees well with the Born approximation result~\cite{Bohn:2014},
\begin{align}
    \label{eq:Born-length}
    a^{\mathrm{Born}}_{02} = -\frac{a_d}{3\sqrt{5}} \,,
\end{align}
except at the $s$-wave resonance. Exactly on resonance, the value of $\text{Re}\,a_{02}$ goes to zero, and the width of the dip in $\text{Re}\,a_{02}$ is determined by the momentum $k$. Using a modified form of the Born approximation, taking into account the $s$-wave scattering length, we find that the feature in $\text{Re}\,a_{02}$ has a universal shape,
\begin{equation}
    \label{eq:dipolar-length}
    \text{Re}\,a_{02}=\frac{a^{\mathrm{Born}}_{02}}{1+k^2 a^2} \,, \quad k a_d \ll 1 \,.
\end{equation}
Moreover, the corresponding cross section~\cite{Bohn:2014},
\begin{equation}
    \sigma_{02}(k) = \frac{\pi}{k^2} |T^0_{02}(k)|^2 \,,
\end{equation}
follows also a universal form given by
\begin{equation}
    \label{eq:cross-section}
    \sigma_{02}(k) = \frac{4 \pi (a^{\mathrm{Born}}_{02})^2}{1+k^2 a^2} \,, \quad k a_d \ll 1 \,.
\end{equation}
However, for larger momentum, we find deviations from the universal behavior, see Fig.~\ref{fig:2}(c). For details on the derivation of the universal expressions for $a_{02}$ and $\sigma_{02}$ close to the $s$-wave resonance, we refer to Appendix~\ref{app:dipolar-dip}.

\subsection{Phase shifts and effective-range expansion} \label{subsec:phase-shift}

Beyond scattering lengths, the low-energy physics is encoded in the momentum-dependent scattering phase shifts $\delta^m_l(k)$, which are defined by
\begin{equation}
    S^m_{ll}(k) = e^{2i \delta^m_l(k)} \,.
\end{equation}
The energy dependence of the dipolar $s$-wave phase shift deviates strongly from the usual effective-range expansion for contact interactions. Because the dipolar interaction has no direct matrix element in the $s$-wave channel, the effective-range expansion also differs from the result of an effective isotropic $r^{-3}$ interaction. Indeed, one can show that the $s$-wave channel asymptotically experiences an effective $r^{-4}$ potential through coupling between the $l=0$ and $l=2$ angular-momentum channels~\cite{Bohn:2009,PhysRevA.85.020702,PhysRevA.98.062502}. The effective-range expansion is therefore reminiscent of the $r^{-4}$ expansion~\cite{omalley:1961}, although the numerical factors differ. The modified effective-range expansion for dipolar potentials was computed by Fabrikant~\cite{Fabrikant:1981,Fabrikant:1984}, originally in the context of charge-quadrupole scattering. For $l=m=0$, it is given by
\begin{equation}\label{eq:expansion_fabrikant}
    k^{-1}\tan\delta^0_0(k) = -a-\frac{76\pi}{525} a^2_d k -\frac{8a}{15} a^2_d k^2 \ln|k a_d| + \mathcal{O}(k^2) \,.
\end{equation}
Here, the terms proportional to $k$ and $k^2\ln|ka_d|$ are universal and arise from the anisotropic long-range tail of the dipolar interaction, while terms of order $k^2$ and higher encode non-universal short-range physics. We verified that this form is consistent with our numerical results.

Note that for the dipolar polaron we are interested in the regime where $k_Fa_d \approx 1$. This implies that the finite energy corrections from Eq.~\eqref{eq:expansion_fabrikant} are substantial. Both the universal terms and the non-universal contributions are important, because the typical effective range is of the order of $a_d$. This is different for contact interactions, where the effective range (except at narrow Feshbach resonances~\cite{Massignan:2012_narrow}) is typically comparable to the van der Waals length, which is orders of magnitude smaller than the interparticle distance.

For the $p$-wave channel, a direct Born contribution arises from the diagonal matrix element $C_{11}^0=2/5$, see also~\cite{PhysRevA.98.062502}. The low-energy expansion of the $p$-wave phase shift ($l=1$, $m=0$) therefore takes the form
\begin{align}\label{eq:p-wave-expansion}
    k^{-1}\tan\delta^0_1(k) = \frac{2}{5}a_d + \mathcal{O}(k) \,.
\end{align}
The subleading terms follow a similar modified effective-range expansion as the $s$-wave expression in Eq.~\eqref{eq:expansion_fabrikant}, with universal long-range coefficients and non-universal higher-order terms~\cite{Fabrikant:1984}. In contrast to the short-range case, where the leading $p$-wave contribution scales as $k^2$ and defines a scattering volume, the dipolar tail produces a finite $p$-wave scattering length set by $a_d$.

These energy-dependent, multichannel phase shifts serve as direct input for the polaron properties, which we discuss next.

\section{Many-body properties} \label{sec:many-body-problem}

In this Section, we investigate the many-body response of a non-interacting Fermi gas coupled to a static dipolar impurity. Owing to the infinite impurity mass and the absence of interactions in the bath, the Hamiltonian remains quadratic in fermionic operators and the dynamics can be solved numerically exactly. The absence of impurity recoil allows to establish a direct connection between the few-body scattering properties from Sec.~\ref{sec:two-body-problem} and many-body observables, including the polaron energy, spectral response, and Anderson’s orthogonality catastrophe.

We consider a single static impurity in a three-dimensional Fermi sea, coupled to the bath fermions via dipole–dipole interactions, while interactions among the bath particles are neglected. This corresponds to a local dipolar scattering potential for the non-interacting Fermi sea. The Hamiltonian is given by
\begin{align} \label{eq:real-space-hamiltonian}
    \hat{H} = \int_{\bm{r}} \hat{\psi}^\dagger(\bm{r})
    \left(-\frac{\nabla^2}{2m}\right) \hat{\psi}(\bm{r}) + \int_{\bm{r}} \hat{\psi}^\dagger(\bm{r}) V(\bm{r})
    \hat{\psi}(\bm{r}) \,,
\end{align}
where $\hat{\psi}^{\dagger}(\bm{r})$ creates a fermion of mass $m$ at position $\bm{r}$, and the impurity is located at the origin.

Neglecting interactions within the Fermi sea is well motivated if the bath fermions have a weaker (effective) dipole moment than the impurity, such that $|k_F a_d^{(\mathrm{bath})}|\ll1$. In this regime, dipolar interactions in a single-component Fermi gas primarily lead to deformations of the Fermi surface~\cite{Nakano:2024the}, which have only minor quantitative effects on the many-body properties addressed here. Moreover, mixtures of ultracold molecules in different vibrational states can provide an experimental realization of this regime, where double microwave shielding can be used to cancel dipolar interactions in the bath, while maintaining dipolar impurity-bath interactions~\cite{jozwiak:2026,Yang:2026}.

\subsection{Dipolar polaron energy} \label{subsec:energy}

The exact ground-state energy of a static impurity in a Fermi sea is given by the generalized Fumi’s theorem, which relates the polaron energy $E_{\mathrm{pol}}$ directly to the two-body scattering phase shifts~\cite{Fumi:1955}. In the presence of bound states with energies $E^{(i)}_b<0$, the polaron energy is given by
\begin{equation}\label{eq:fumi-polaron}
    E_{\mathrm{pol}} = -\frac{1}{\pi} \int_0^{E_F} dE \sum_{nm} \tilde{\delta}^m_{n}(E) + \sum_i E^{(i)}_b \,.
\end{equation}
Here, $m$ denotes the conserved magnetic quantum number and $n$ labels the eigenchannels of the $S$-matrix within a given $m$-sector. The eigenphase shifts $\tilde{\delta}_n^m$ are defined by
\begin{equation}
\tilde{S}_{nn}^m(k)=e^{2i\tilde{\delta}_n^m(k)} \,,
\end{equation}
where $\tilde{S} = P^{-1} S P$ is the diagonalized $S$-matrix, and the columns of $P$ are the eigenvectors of $S$. The sum of eigenphase shifts entering Fumi's theorem is evaluated from the $T$-matrix using the relation given in Eq.~\eqref{eq:generalized-Fumi-shift}. The second term in Eq.~\eqref{eq:fumi-polaron} accounts for bound states, and contributes once the even- or odd-wave bound states become occupied.

\begin{figure}[t]
    \centering
    \includegraphics[width=\linewidth]{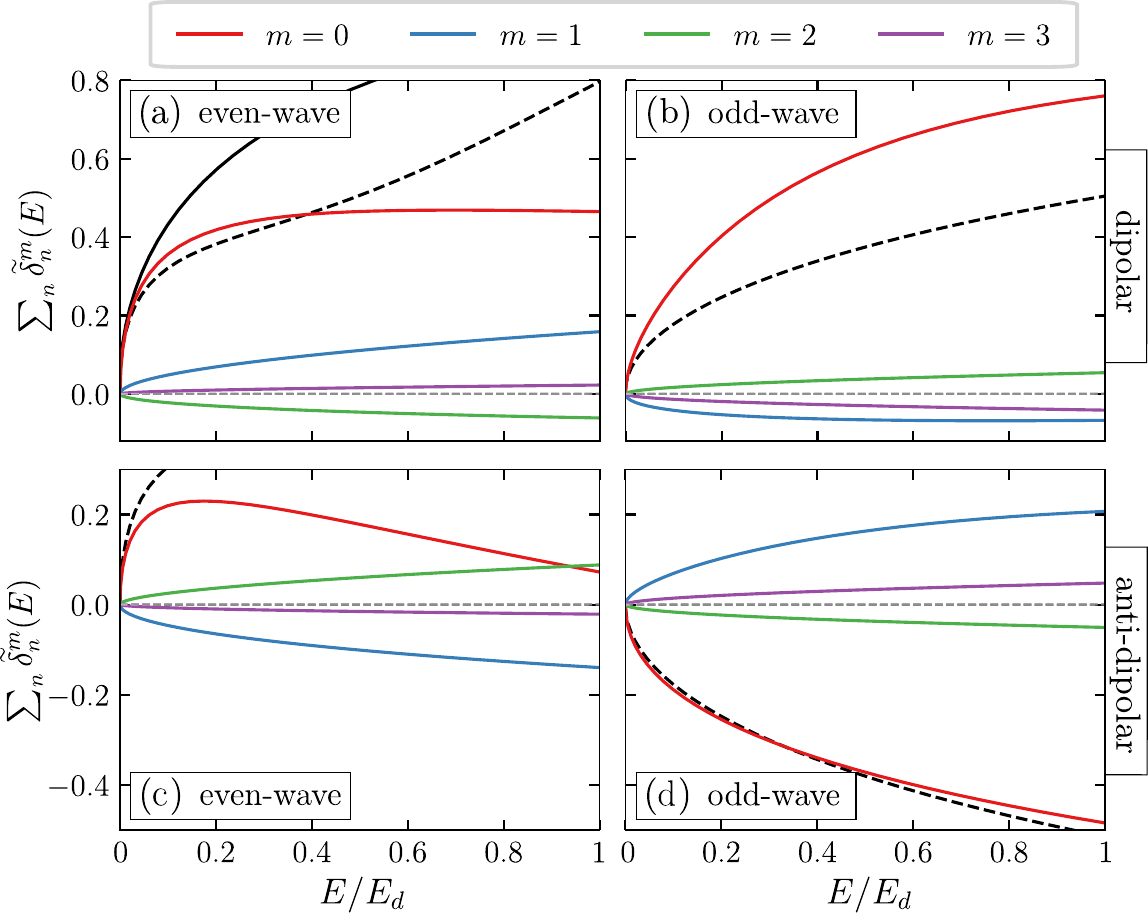}
    \caption{Sum of eigenphase shifts $\sum_n \tilde{\delta}^m_n(E)$ as a function of energy $E/E_d$ at fixed scattering length $a/|a_d| = -1$, for different magnetic quantum numbers $m$. (a,c)~Even-wave contributions for dipolar and anti-dipolar interactions, respectively. (b,d)~The corresponding odd-wave contributions. The black dashed lines show $\delta_{0}^{0}(E)$ and $\delta_{1}^{0}(E)$ from the expansions in Eqs.~\eqref{eq:expansion_fabrikant}     and~\eqref{eq:p-wave-expansion}, truncated at the order shown. The black solid line in panel (a) shows the contact-interaction result for $\delta_{0}^{0}(E)$.}
    \label{fig:3}
\end{figure}

The sum of eigenphase shifts is shown in Fig.~\ref{fig:3} for both dipolar and anti-dipolar interactions at a representative $a/|a_d| = -1$, for different magnetic quantum numbers $m$. The even-wave sector is dominated at low energies by the $s$-wave channel. Its phase shift rises steeply near threshold, but then saturates already at small energies $E/E_d\sim0.2$. This early saturation is a signature of large finite-range corrections, as seen from comparison with the contact-interaction result (black solid line). The odd-wave sector is governed by the $p$-wave channel and acquires equally strong finite-range corrections. In both sectors, the leading-order result derived by Fabrikant~\cite{Fabrikant:1981,Fabrikant:1984} (black dashed lines) captures only the behavior at very low energies, and higher-order terms contribute already at small energies. Comparing dipolar and anti-dipolar interactions, the phase shifts change sign in every $m$-sector, except the even-wave $m=0$ channel, whose leading contribution arises at second order and therefore does not change sign.

\begin{figure}[t]
    \centering
    \includegraphics[width=\linewidth]{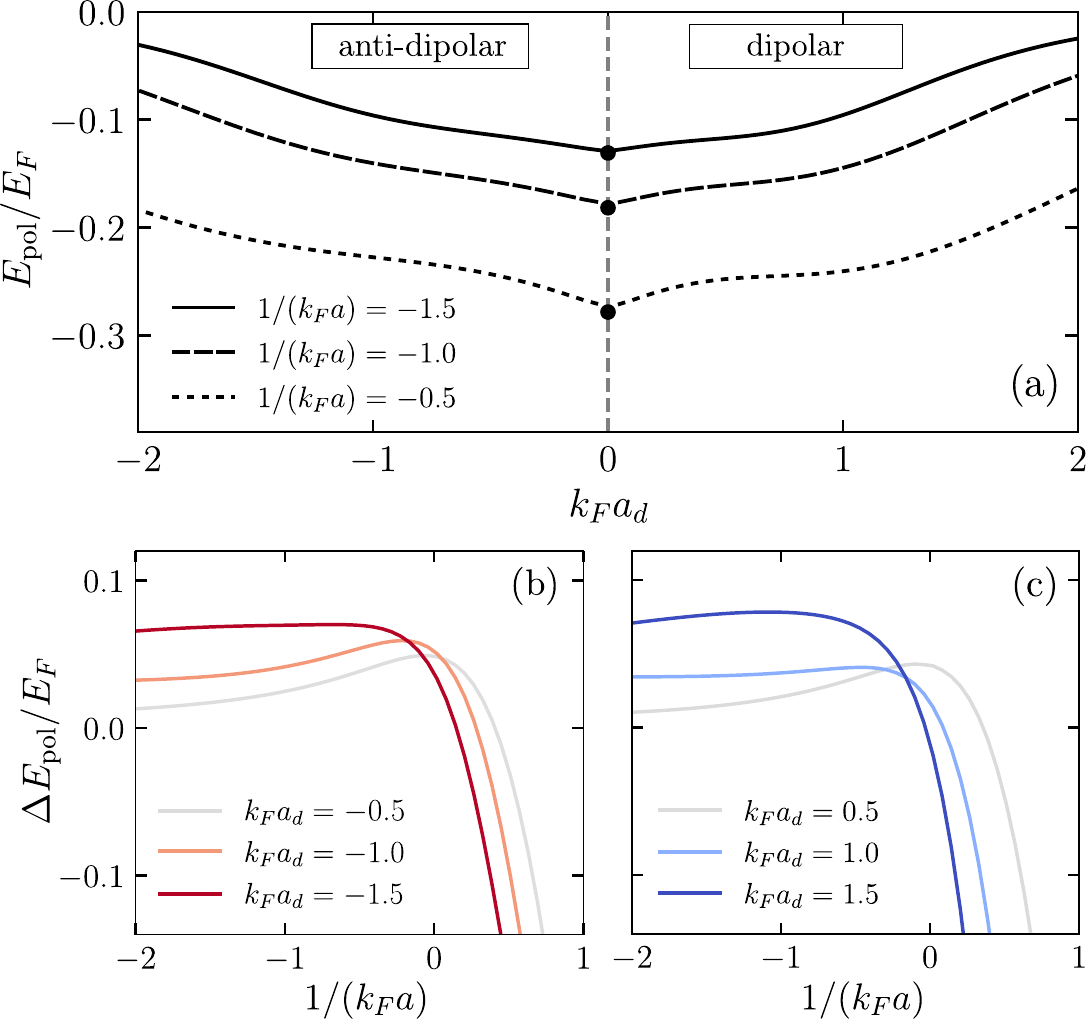}
    \caption{Dipolar polaron energy from Fumi's theorem. (a)~Polaron energy $E_{\mathrm{pol}}$ in units of $E_F$ as a function of dipole strength $k_Fa_d$, for three inverse scattering lengths $1/(k_Fa)$. The black points denote the results from contact interactions. (b,c)~Relative polaron energy $\Delta E_{\mathrm{pol}}(a_d)=E_{\mathrm{pol}}(a_d)-E_{\mathrm{pol}}(0)$ as a function of $1/(k_Fa)$ for several dipole strengths $k_Fa_d$, on the anti-dipolar (b) and dipolar (c) side.}
    \label{fig:4}
\end{figure}

The resulting polaron energy $E_\mathrm{pol}$ is shown in Fig.~\ref{fig:4}(a) as a function of dipole strength $k_F a_d$ for three inverse scattering lengths $1/(k_F a) = -1.5,\,-1.0,\,-0.5$. For a fixed negative scattering length, the dipolar polaron energy shifts monotonically toward zero with increasing $k_F|a_d|$ in a way that is remarkably similar for both dipolar and anti-dipolar interactions. This repulsive energy shift mostly follows from the suppression of dipolar phase shifts relative to the contact-interaction result at finite energies, together with small contributions from higher magnetic quantum numbers.

To further isolate the role of dipolar interactions, we define the relative polaron energy
\begin{equation} \label{eq:relative-energy}
    \Delta E_{\mathrm{pol}}(a_d) = E_{\mathrm{pol}}(a_d)-E_{\mathrm{pol}}(a_d=0) \,,
\end{equation}
shown in Fig.~\ref{fig:4}(b,c) as a function of $1/(k_F a)$ for several $k_F a_d$. Dipolar interactions produce a pronounced and non-monotonic modification of the polaron energy, directly inherited from the energy dependence of the phase shifts. As seen in Fig.~\ref{fig:4}(a), the polaron energy obtains a repulsive shift with respect to the contact-interacting result for negative scattering lengths, but close to resonance the correction changes sign and becomes attractive. This sign change resembles the effect of a large positive effective range in dipolar-induced $s$-wave resonances~\cite{PhysRevA.85.020702}. The non-trivial effect of the energy-dependent scattering shows that contact and dipolar contributions do not enter additively in the fermionic system. This differs from the Gross-Pitaevskii description of bosonic dipolar systems, where they appear as independent terms in the mean-field energy functional~\cite{Chomaz:2023}.

\subsection{Dipolar polaron spectrum} \label{subsec:spectra}

\begin{figure}[b]
    \centering
    \includegraphics[width=\linewidth]{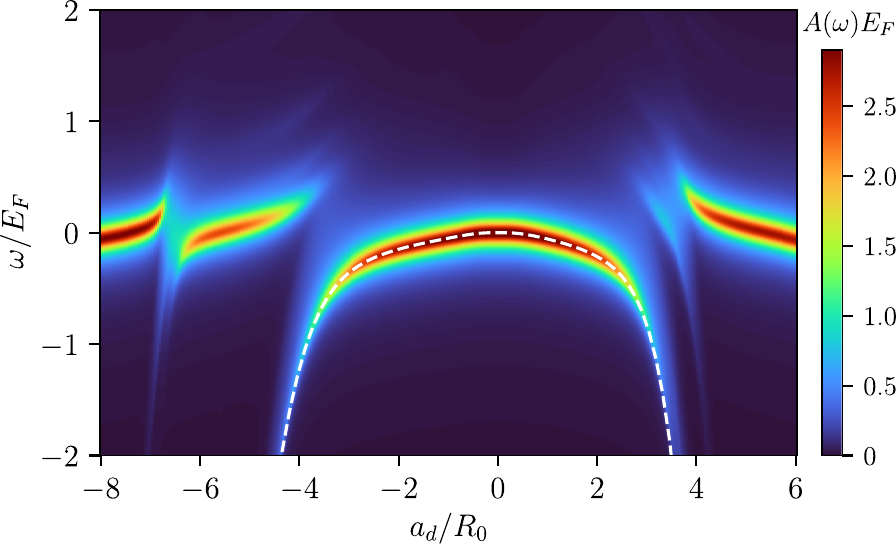}
    \caption{Impurity spectrum $A(\omega)E_F$ at fixed $k_Fa_d=1$ as a function of the inverse short-range cutoff $a_d/R_0$, which tunes the depth of the model potential and hence the bound-state content [see Fig.~\ref{fig:2}(a)]. The polaron energy from Fumi's theorem, Eq.~\eqref{eq:fumi-polaron}, is overlaid as a white dashed line.}
    \label{fig:5}
\end{figure}
\begin{figure*}[t]
    \centering
    \includegraphics[width=\linewidth]{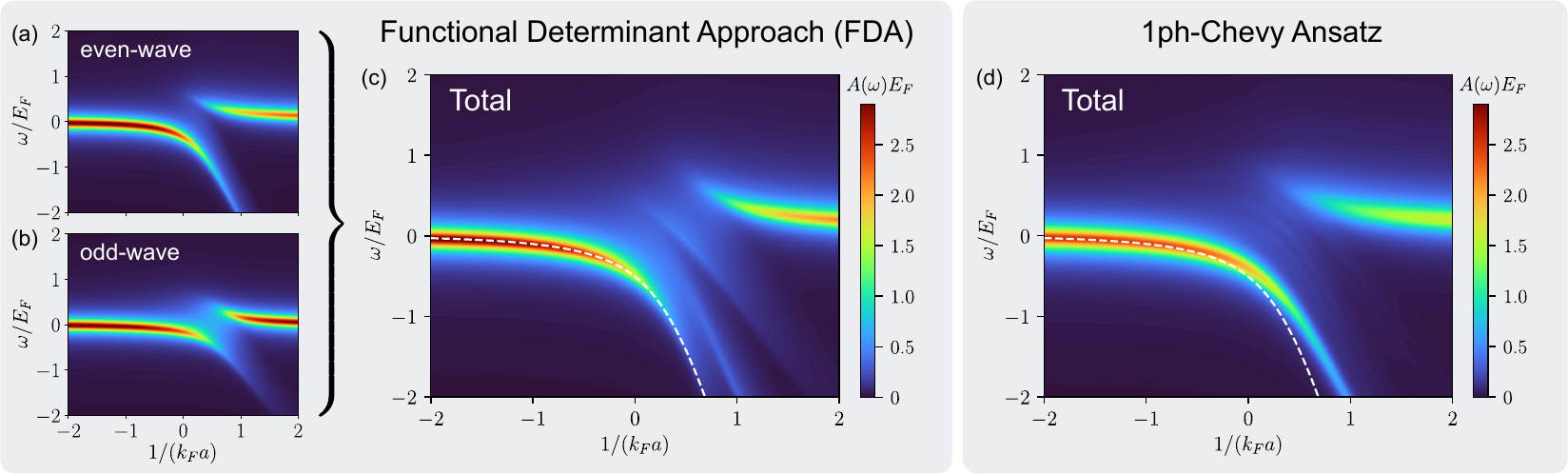}
    \caption{Convolution structure of the FDA spectrum and comparison to the Chevy ansatz at $k_F a_d=1.5$. (a)~Even-wave and (b)~odd-wave contributions to the impurity spectrum, computed separately within the FDA. (c)~Total FDA spectrum, obtained as their convolution, Eq.~\eqref{eq:spectra_convolution}. (d)~Total spectrum from the 1ph-Chevy ansatz, which misses the low-lying ground-state branch arising from the simultaneous occupation of both bound states. The polaron energy from Fumi's theorem, Eq.~\eqref{eq:fumi-polaron}, is overlaid as a white dashed line.}
    \label{fig:6}
\end{figure*}

Beyond the ground-state energy, we compute the full impurity spectral function using the functional determinant approach (FDA)~\cite{Gievers2024,Sous2020,Schmidt2018}. The normalized impurity spectral function is given by
\begin{equation} \label{eq:Aomega}
    A(\omega) = \frac{1}{\pi}\,\text{Re}\,\int_0^{\infty} dt\, e^{i \omega t}\, S(t) \,,
\end{equation}
where $S(t)$ is the so-called Loschmidt echo,
\begin{equation} \label{eq:s_t}
    S(t) = \text{Tr}\left[ e^{i \hat{H}_0 t} e^{-i \hat{H} t} \hat{\rho} \right] \,.
\end{equation}
Here, $\hat{H}_0$ and $\hat{H}$ denote the Hamiltonian in absence and presence of the impurity, respectively. $\hat{\rho}=e^{-\beta(\hat{H}_0-\mu\hat{N})}/Z$ is the density matrix of the free Fermi gas at inverse temperature $\beta=1/T$ and chemical potential $\mu$, and $Z$ is the partition function of the Fermi gas. The Loschmidt echo can be measured experimentally via Ramsey spectroscopy~\cite{Cetina:2016}, and at $T=0$ it equals the overlap between the time-evolved interacting and non-interacting many-body wave functions~\cite{Goussev:2012}.

Owing to the quadratic nature of $\hat{H}_0$ and $\hat{H}$, the overlap~\eqref{eq:s_t} can be evaluated exactly using the FDA,
\begin{equation}
    S(t) = \det\left[1-n_{\mathrm{F}}(\hat{h}_0)+n_{\mathrm{F}}(\hat{h}_0) e^{i \hat{h}_0 t} e^{-i \hat{h} t}\right] \,.
\end{equation}
The determinant is calculated in the single-particle Hilbert space, with $\hat{h}_0$ and $\hat{h}$ being the one-body counterparts of $\hat{H}_0$ and $\hat{H}$, respectively. Moreover, $n_{\mathrm{F}}(\hat{h}_0)=1/(e^{\beta(\hat{h}_0-\mu)}+1)$ is the Fermi-Dirac distribution of the non-interacting single-particle orbitals of $\hat{h}_0$. For details on the numerical implementation, see Appendix~\ref{app:FDA}.

In Fig.~\ref{fig:5}, we show the dimensionless impurity spectrum $A(\omega)E_F$ at $k_F a_d=1$ as a function of $a_d/R_0$. Mirroring the bound-state hierarchy of the two-body problem in Fig.~\ref{fig:2}(a), both even- and odd-wave bound states appear as distinct branches in the impurity spectrum. On the anti-dipolar side ($a_d/R_0<0$), the even- and odd-wave bound states are visibly more separated, and the doubly degenerate $|m|=1$ odd-wave bound state leaves a similar trace in the spectrum as the closely spaced even- and odd-wave bound states on the dipolar side ($a_d/R_0>0$). We discuss the origin of this structure in more detail below. When multiple bound states are present, the many-body ground state corresponds to their simultaneous occupation. This is a feature with no analogue in contact-interacting systems, where only a single bound state exists. The polaron energy from Fumi's theorem (white dashed line) agrees with the ground-state branch in the polaron spectrum.

The total polaron spectrum follows from the convolution of spectra in different symmetry sectors. In the static-impurity limit, the symmetry of the dipole potential decouples both different magnetic quantum numbers and even- and odd-parity sectors, such that the total Loschmidt echo factorizes into a product,
\begin{equation}
    S(t) = S_{\text{even}}(t) \cdot S_{\text{odd}}(t) \,,
\end{equation}
where $S_{\text{even/odd}}(t)=\prod_m S^m_{\text{even/odd}}(t)$ is a product over all magnetic quantum numbers $m$. The total impurity spectrum is therefore a convolution of even- and odd-wave contributions,
\begin{equation} \label{eq:spectra_convolution}
    A(\omega) = \{A_{\text{even}} \ast A_{\text{odd}}\}(\omega) \,.
\end{equation}
This gives a direct physical interpretation of the spectral features in terms of the underlying scattering channels. Each sector carries its own bound-state and continuum structure, and the convolution combines them, such that a peak in $A_{\text{even}}$ and a peak in $A_{\text{odd}}$ produce a joint feature at the sum of their energies.

The convolution structure is shown in Fig.~\ref{fig:6}(a-c) at $k_Fa_d=1.5$ for dipolar interactions, now as a function of $1/(k_Fa)$. Panels (a) and (b) show the even- and odd-wave spectra computed separately, and panel (c) their convolution, which reproduces the full FDA result and makes the joint even-odd features explicit. For anti-dipolar interactions ($k_Fa_d<0$), there are two degenerate bound states in the odd-wave spectrum for $m=\pm 1$, which lead to a similar total spectrum.

For comparison, in Fig.~\ref{fig:6}(d) we also show the spectrum from the one particle-hole (1ph) Chevy ansatz~\cite{PhysRevA.74.063628}, a controlled variational approach widely used in polaron studies~\cite{PhysRevA.103.033324,PhysRevLett.109.075302}. In the impurity frame, the 1ph-Chevy ansatz reads~\cite{PhysRevA.103.033324}
\begin{align} \label{eq:LLP-chevy-ansatz}
    |\Psi\rangle = \alpha_0\, |\mathrm{FS}\rangle + \sum_{\mathbf{k}>,\mathbf{q}<} \alpha_{\mathbf{kq}}\, \hat{c}^{\dagger}_{\mathbf{k}}\hat{c}_{\mathbf{q}}|\mathrm{FS}\rangle \,,
\end{align}
where $c^{\dagger}_{\mathbf{k}}$ ($c_{\mathbf{q}}$) create (annihilate) fermions with momenta $\mathbf{k}>$ ($\mathbf{q}<$) above (below) the Fermi momentum $k_F$, and $|\mathrm{FS}\rangle=\prod_{\mathbf{q}<} c^{\dagger}_{\mathbf{q}} |0\rangle$ is the Fermi sea. Here, $\alpha_0$ and $\alpha_{\mathbf{kq}}$ are variational parameters, which are determined by minimizing the energy $E=\langle\Psi|\hat{H}|\Psi\rangle/\langle\Psi|\Psi\rangle$, equivalent to diagonalizing the Hamiltonian in a subspace. For more details, we refer to Appendix~\ref{app:chevy-ansatz}.

\begin{figure*}[t]
    \centering
    \includegraphics[width=\textwidth]{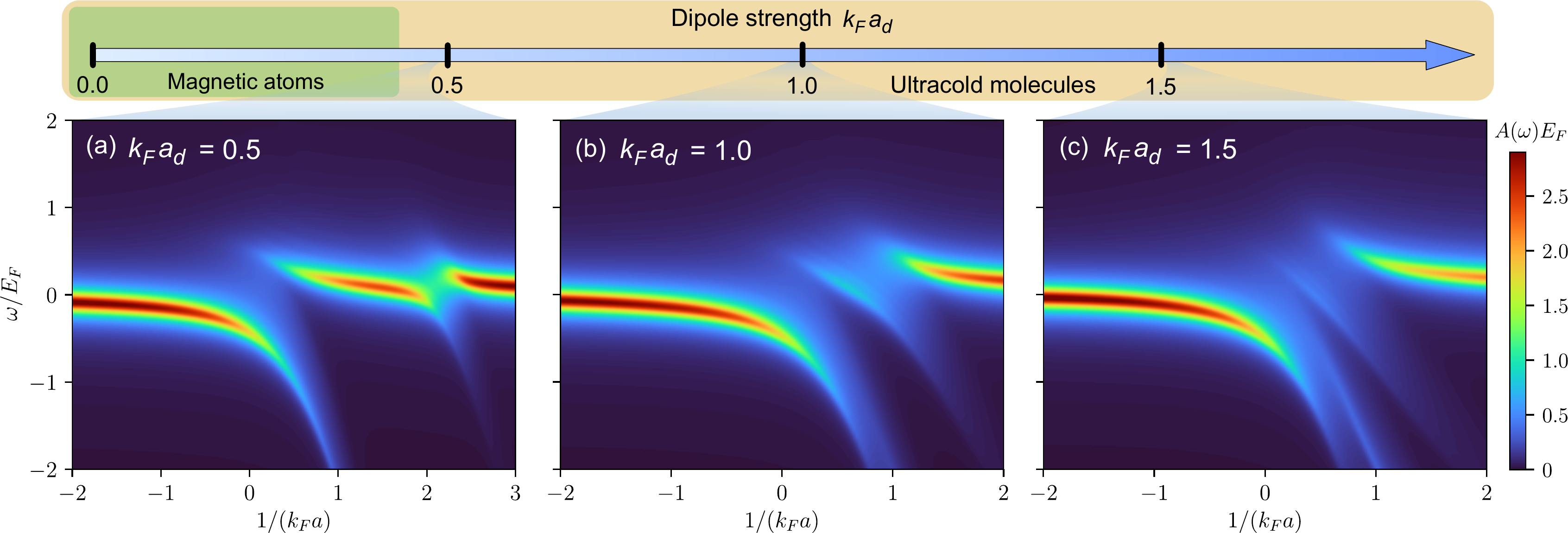}
    \caption{Impurity spectral function $A(\omega)E_F$ as a function of inverse scattering length $1/(k_F a)$, for three dipole strengths (a)~$k_F a_d = 0.5$, (b)~$k_F a_d = 1.0$, and (c)~$k_F a_d = 1.5$. At positive scattering length, the spectrum splits into an even-wave and an odd-wave bound-state branch, whose specific lineshape follows from the convolution structure of Eq.~\eqref{eq:spectra_convolution}. The branch separation shrinks with increasing $k_F a_d$. Panel (a) lies just above the dipole strengths of magnetic dysprosium atoms, while panels (b,c) are accessible with microwave-shielded polar molecules.}
    \label{fig:7}
\end{figure*}

While the Chevy ansatz captures the overall polaron energy shift reasonably well at negative scattering length, it fails to resolve the spectral structure at positive scattering length. Since the Chevy ansatz contains at most one particle-hole excitation, it cannot capture the ground state in the regime where the even- and odd-wave bound states are simultaneously occupied. Instead, the Chevy ansatz couples the even and odd angular-momentum channels through the $\alpha_0$ coefficient, leading to a ground-state wave function corresponding to a superposition of the two bound states, see Appendix~\ref{app:chevy-ansatz} for a detailed discussion. These findings indicate that higher-order particle–hole excitations are essential for a quantitatively accurate description of dipolar polarons, and that the Chevy approach should be used with caution in the strongly dipolar regime. While this hybridization is an artifact of the ansatz for static impurities, this effect could become relevant for finite-mass impurities.

Figure~\ref{fig:7} shows the impurity spectral function $A(\omega)E_F$ for three dipole strengths $k_Fa_d$. The even- and odd-wave bound states appear at different values of $1/(k_Fa)$, giving rise to two separate branches in the spectrum. In panel (a), these branches are clearly resolved. With increasing $k_Fa_d$, the two bound states approach in $1/(k_Fa)$~\footnote{The odd-wave onset sits at $1/(k_Fa) = (a_d/a)_c/(k_Fa_d)$, so it approaches resonance as $k_Fa_d$ grows.} and move closer together in the spectrum, as seen in panels (b) and (c). In this regime, the convolution of the two branches produces the doubly occupied ground state and a richer spectral structure.

\begin{figure}[b]
    \centering
    \includegraphics[width=\linewidth]{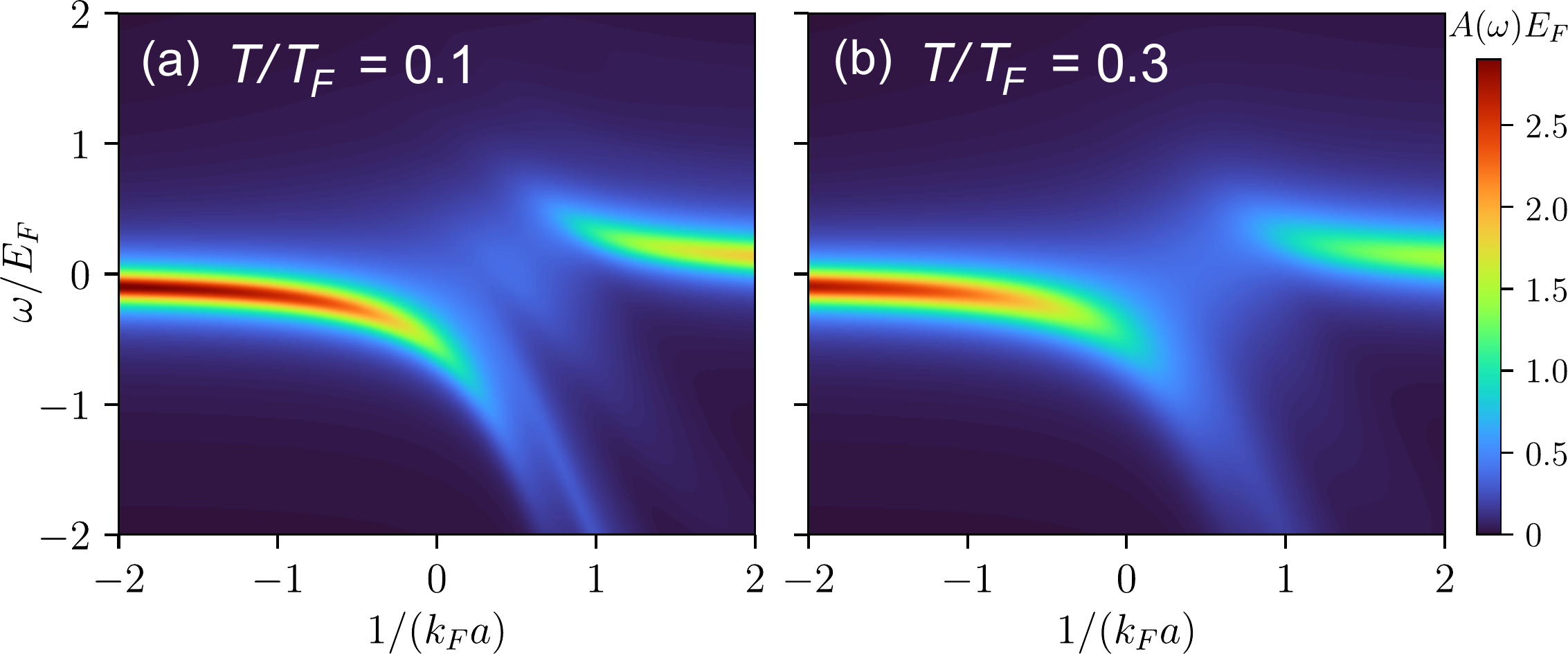}
    \caption{Finite-temperature impurity spectrum $A(\omega)E_F$ at $k_F a_d=1.5$ for (a)~$T/T_F=0.1$ and (b)~$T/T_F=0.3$. The even- and odd-wave branches remain resolved at $T/T_F=0.1$, but become barely distinguishable at $T/T_F=0.3$.}
    \label{fig:8}
\end{figure}

The entire shown parameter range is experimentally accessible. Microwave-shielded polar molecules can reach dipole lengths as large as $a_d \gtrsim 10^4\,a_0$ depending on the species~\cite{schindewolf:2026,Karman2025DMS,Deng:2025fjv}, making $k_Fa_d \approx 1$ and beyond achievable at realistic densities. The spectral function in Eq.~\eqref{eq:Aomega} is accessible via injection or Ramsey spectroscopy~\cite{Cetina:2016}, where two internal states of the impurity serve as the interacting and non-interacting branches of the Ramsey sequence. Ultracold molecules offer a wide variety of internal states that can be used as reference states, although collisional losses and spectroscopic constraints must be considered~\cite{Yang:2026}.

Magnetic dysprosium atoms, with a dipole length of $a_d\approx200\,a_0$~\footnote{Our convention $a_d = m_r d^2/\hbar^2$ gives $a_d = m\mu_0\mu^2/(8\pi\hbar^2)$ with $d^2 = \mu_0\mu^2/(4\pi)$. This differs by a factor of $3/2$ from the bosonic convention $a_{\rm dd} = m\mu_0\mu^2/(12\pi\hbar^2)$~\cite{Chomaz:2023}, under which dysprosium has $a_{\rm dd}\approx130\,a_0$.}, reach $k_Fa_d\approx0.1$--$0.4$ at realistic densities~\cite{PhysRevLett.108.215301,soave:2022}. Since Feshbach resonances in magnetic atoms are typically narrow, the polaron spectrum is not expected to match precisely the resonance structure studied here. The relative positions of possible even- and odd-wave bound states would instead be set by the positions of the respective Feshbach resonances. Moreover, magnetic atoms are typically only stable against collisional dipolar relaxation in their lowest spin state~\cite{Chomaz:2023}, limiting the choice of spectroscopic reference states.

The above spectra are computed at zero temperature. To assess their robustness under realistic experimental conditions, we also perform FDA calculations at finite temperature, shown in Fig.~\ref{fig:8}. Raising the temperature broadens the spectral lines and slightly shifts the peak positions, but leaves the even-odd branch structure intact as long as $T/T_F\lesssim0.3$. The temperatures required to resolve the predicted features are therefore within reach of realistic experiments~\cite{demarco:2019,Wang:2024}. Only above this scale does thermal broadening obscure the distinction between the bound-state branches.

\subsection{Orthogonality catastrophe} \label{subsec:orthogonality}

Finally, we investigate the Anderson orthogonality catastrophe (OC) in the presence of dipolar interactions. Originally formulated for static impurities with short-range potentials~\cite{Anderson:1967}, the OC predicts that the overlap between the interacting and non-interacting many-body ground states vanishes in the thermodynamic limit. The quasiparticle weight $Z$ exhibits a power-law decay,
\begin{equation}
    Z = |\langle\mathrm{FS}|\Psi\rangle|^2 \sim N^{-\alpha} \,,
\end{equation}
where $N$ is the fermion number, and the exponent $\alpha$ is determined by the scattering phase shifts~\cite{Nozieres:1969,Combescot:1971},
\begin{equation} \label{eq:exponent}
    \alpha = \sum_{nm} \left( \frac{\tilde{\delta}^m_n(E_F)}{\pi} \right)^2 \,.
\end{equation}
For central potentials, the eigenchannels $n$ reduce to usual partial waves $l$, and the $m$-sum gives a degeneracy factor, recovering the familiar $\alpha=\sum_l(2l+1)[\delta_l(E_F)/\pi]^2$ \cite{Anderson:1967}. For dipolar interactions, the anisotropy lifts the $m$-degeneracy and mixes partial waves, so the sum runs over all magnetic quantum numbers $m$ and all diagonalized eigenchannels $n$ within each parity sector. Because even and odd sectors decouple, the exponent splits additively,
\begin{equation} \label{eq:oc-split}
    \alpha = \alpha_{\text{even}} + \alpha_{\text{odd}} \,.
\end{equation}
This is the OC analogue of the spectral convolution in Eq.~\eqref{eq:spectra_convolution}, where the same decoupling that convolves the spectrum adds the OC exponents. However, unlike the polaron energy in Eq.~\eqref{eq:fumi-polaron}, where phase shifts of opposite sign can lead to cancellations, the exponent $\alpha$ is a sum of positive-definite contributions.

\begin{figure}[t]
    \centering
    \includegraphics[width=\linewidth]{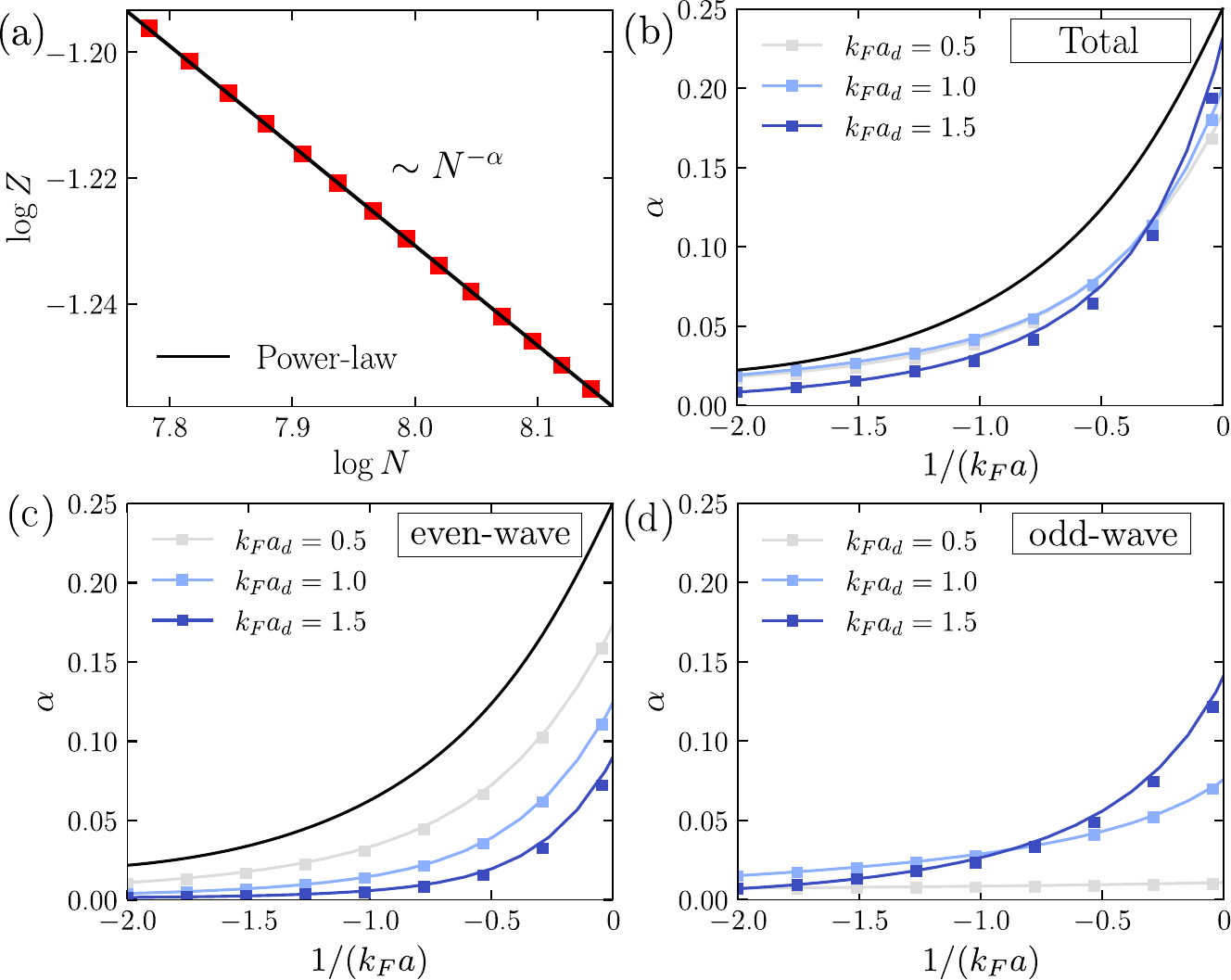}
    \caption{Anderson orthogonality catastrophe for a dipolar impurity. (a)~Power-law decay of the quasiparticle weight $Z\sim N^{-\alpha}$ as a function of system size $N$. (b)~Exponent $\alpha$ as a function of inverse scattering length $1/(k_F a)$ for several dipole strengths $k_F a_d$, compared with the analytic prediction of Eq.~\eqref{eq:exponent} using numerical phase shifts (solid lines). (c)~Even-wave and (d)~odd-wave contributions to the total exponent. The black solid line in panels (b) and (c) shows the contact-interaction result.}
    \label{fig:9}
\end{figure}

Figure~\ref{fig:9}(a,b) show the power-law decay of $Z$ as a function of system size $N$, and the extracted exponent $\alpha$ as a function of inverse scattering length $1/(k_F a)$ for several dipole strengths $k_F a_d$. The exponent extracted from the finite-size scaling (squares) agrees well with the analytic expression in Eq.~\eqref{eq:exponent}, evaluated using the numerical scattering phase shifts (solid lines). We find that increasing dipole strength leads to a reduction of $\alpha$, indicating a slower decay of the quasiparticle weight. This behavior can be understood from the modified effective-range expansion, similar to what we found in Sec.~\ref{subsec:energy}. As $k_Fa_d$ grows while keeping $k_Fa$ fixed, the phase shift at the Fermi energy is suppressed, thereby reducing $\alpha$. Quantitatively, the dominant contributions arise from the $s$- and $p$-wave channels, while higher partial waves are strongly suppressed. Resolving the two parity sectors [Fig.~\ref{fig:9}(c,d)] shows their different behavior. The even-wave contribution decreases monotonically with $k_Fa_d$, while the odd-wave contribution depends non-trivially on dipole strength and scattering length.

The reduction of $\alpha$ has a direct experimental signature. The exponent controls the edge singularity of the spectral function, $A(\omega)\sim|\omega-\omega_0|^{\alpha-1}$, and the long-time decay of the Loschmidt echo, $|S(t)|\sim t^{-\alpha}$. A smaller $\alpha$ therefore means slower decoherence and sharper quasiparticle peak in the spectra of Sec.~\ref{subsec:spectra}. Thus, dipolar polarons retain coherence longer than their contact counterparts. Our results connect smoothly to the known contact limit~\cite{Chen:2025} and show how long-range anisotropic forces reshape the orthogonality catastrophe in dipolar Fermi systems.

%

\section{Conclusion \& Outlook} 
\label{sec:Conclusion}

We have studied a static dipolar impurity in a non-interacting Fermi gas and connected its two-body scattering physics to the resulting polaron properties. We characterized the dipolar bound-state structure and the energy-dependent phase shifts, and showed through Fumi's theorem how these phase shifts enter the polaron energy. This energy varies non-trivially with dipole strength, revealing that dipolar and contact interactions do not enter additively. Computing the full spectrum with the functional determinant approach, we found that it factorizes into a convolution of even- and odd-wave contributions, giving a ground state with simultaneously occupied even- and odd-wave bound states. The separation between these bound states is directly tunable via the dipole length. Comparison with the 1ph-Chevy ansatz revealed that a single particle–hole excitation cannot capture the true ground state, as it cannot describe the simultaneous occupation of both bound states. Finally, we showed that long-range interactions weaken Anderson's orthogonality catastrophe, reducing the power-law decay of the quasiparticle weight, with the exponent splitting additively over parity sectors.

These predictions are accessible in realistic experiments with ultracold molecules~\cite{jozwiak:2026,Yang:2026}. The cleanest signature is the tunable separation between the even- and odd-wave bound-state branches. This is observable in the absorption spectrum of microwave-shielded polar molecules at temperatures $T/T_F \lesssim 0.3$, where short-range and dipolar interactions can be tuned independently.

A natural next step is the extension to mobile impurities, where the parity sectors no longer decouple. The parity mixing that appears as an artifact of the Chevy ansatz at infinite mass may then become relevant. A particularly interesting feature at finite impurity mass is the polaron-to-molecule transition~\cite{Punk:2009,PhysRevA.103.063312}, for which the mass-gap description of Ref.~\cite{Chen:2025} provides a promising theoretical route~\cite{Yang:2026}. Beyond the single-impurity limit, our results connect naturally to the phase diagram of unconventional pairing~\cite{Read:2000,Baranov:2002,Gurarie:2005,cooper:2009,levinsen:2011,deng:2023,PhysRevLett.110.045302} in density-imbalanced dipolar Fermi mixtures.

\subsection*{Acknowledgements} \label{sec:Acknowledgments}

We thank M.~Gievers for helpful discussions on the FDA and J.~Bohn for his suggestions on modeling the properties of the dipolar length $a_{02}$ at the $s$-wave scattering resonance. Moreover, we thank T.~Karman, H.~Yang and C.~R.~Cabrera for fruitful discussions. This work is funded by the DFG (German Research Foundation) – Project-ID 273811115 – SFB 1225 ISOQUANT, and under Germany's Excellence Strategy EXC 2181/1 - 390900948 (the Heidelberg STRUCTURES Excellence Cluster). A.C. was funded by an ETH Fellowship.

\vfill

\newpage

\begin{widetext}
\appendix

\section{Dipolar scattering calculations} \label{app:dipolar-scattering}

As described in the main text, using the model potential in Eq.~\eqref{eq:model-potential-real}, the radial wave functions $F_{lm}(r)$ satisfy the coupled-channel Schrödinger equations,
\begin{align}
    -\frac{1}{2}\frac{d^2}{dr^2}F_{lm}(r) + \frac{l(l+1)}{2r^2}F_{lm}(r) - \frac{2g(r)}{r^3} \sum_{l'} C^{m}_{ll'} F_{l'm}(r) = E F_{lm}(r) \,,
\end{align}
where the coupling matrix elements $C^{m}_{ll'}$ are given by~\cite{Bohn:2009}
\begin{align} \label{eq:coupling-matrix-element-SM}
    C^{m}_{ll'} = \langle lm|C_{20}|l'm\rangle = (-1)^m \sqrt{(2l+1)(2l'+1)} 
    \begin{pmatrix}
        l & 2 & l' \\
        -m & 0 & m
    \end{pmatrix}
    \begin{pmatrix}
        l & 2 & l' \\
        0 & 0 & 0
    \end{pmatrix} \,.
\end{align}
expressed in terms of Wigner $3j$ symbols. Since the dipole potential has even parity under $\mathbf{r}\rightarrow-\mathbf{r}$, the coupling matrix $C^{m}_{ll'}$ vanishes between channels of different parity. Even and odd angular-momentum channels therefore decouple completely, and can be treated independently.

For the scattering calculations, the coupled-channel equations are solved numerically using a variable-stepsize Numerov algorithm~\cite{Christianen:2023qig}, propagating from short to large distances and matching to the asymptotic free-particle solutions. The $T$-matrix elements $T^{m}_{ll'}(k)$ are extracted from the asymptotic behavior of the radial wave functions, and the scattering phase shifts are obtained via $S^m_{ll'}(k) = \delta_{ll'} - iT^m_{ll'}(k)$. For all scattering results, we include partial waves up to $l_\mathrm{max} = 30$, well beyond what is needed for convergence.

For the anti-dipolar case ($a_d < 0$), the same equations apply with the sign of the interaction reversed. As discussed in Sec.~\ref{subsec:scattering}, the odd-wave bound state in this case appears in the $|m|=1$ sector rather than $m=0$, reflecting the sign change of the diagonal matrix elements $C^{m}_{11}$.

\section{Functional determinant approach} \label{app:FDA}

The functional determinant approach (FDA) provides an exact method for computing the impurity spectral function for a static impurity coupled to a non-interacting Fermi gas~\cite{Schmidt2018}. It exploits the fact that the Hamiltonian is quadratic in the fermionic operators, so that all many-body quantities reduce to determinants in the single-particle Hilbert space.

We work in a large but finite spherical box of radius $R$, imposing vanishing boundary conditions. The non-interacting single-particle eigenstates are labeled by $|nlm\rangle$,
\begin{align}
	\langle\bm{r}|nlm\rangle = \phi_{nlm}(r,\theta,\phi) = \frac{u_{nl}(r)}{r}\, Y_{lm}(\theta,\phi) \,,
\end{align}
where $u_{nl}(r)=\mathcal{N}_{nl}\, r\, j_l(k_{nl} r)$ is expressed in terms of spherical Bessel functions $j_l(r)$, with $k_{nl}$ determined by the boundary condition $j_l(k_{nl}R) = 0$, and $\mathcal{N}_{nl}$ is a normalization constant~\cite{Gievers2024}. The corresponding non-interacting energies are $E_{nl} = k_{nl}^2/(2m)$.

Since $m$ is a good quantum number, the full Hilbert space decomposes into independent sectors labeled by $m$, and within each sector even and odd parity channels decouple. The interacting single-particle Hamiltonian $\hat{h} = -\nabla^2/(2m) + V(\boldsymbol{r})$ is diagonalized numerically for each magnetic quantum number and parity sector independently. The interacting eigenstates $|\nu\rangle$ with energies $E_\nu$ are obtained by expanding in the non-interacting basis and diagonalizing the resulting matrix. The overlaps between interacting states $|\nu\rangle$ and non-interacting states $|nlm\rangle$ are given by
\begin{align}
	\langle\nu|nlm\rangle = \int d^3r\, \langle\nu|\bm{r}\rangle\langle\bm{r}|nlm\rangle  = \int_0^R dr\, F^{(\nu)}_{lm}(r) u_{nl}(r) \,.
\end{align}
Here, $F^{(\nu)}_{lm}(r)$ are the radial components of the interacting eigenstate $|\nu\rangle$, which are computed numerically using exact diagonalization. The time-dependent Ramsey overlap is given by
\begin{align}
	S(t) = \det\left[1-n_{\mathrm{F}}(\hat{h}_0)+n_{\mathrm{F}}(\hat{h}_0) e^{i \hat{h}_0 t} e^{-i \hat{h} t}\right] \,,
\end{align}
where the determinant is evaluated in the single-particle Hilbert space, and $n_{\mathrm{F}}(\hat{h}_0) = 1/(e^{\beta(\hat{h}_0 - \mu)} + 1)$ is the Fermi-Dirac distribution of the non-interacting system. In the eigenbasis of the non-interacting states, the matrix inside the determinant takes the form
\begin{align}
	&\langle nlm|[1-n_{\mathrm{F}}(\hat{h}_0)+n_{\mathrm{F}}(\hat{h}_0) e^{i \hat{h}_0 t} e^{-i \hat{h} t}]|n'l'm'\rangle \notag \\
	&= \delta_{mm'} \left[ [1-n_{\mathrm{F}}(E_{nl})]\delta_{nn'}\delta_{ll'} + n_{\mathrm{F}}(E_{nl}) e^{i E_{nl} t} \sum_{\nu} \langle nlm|\nu\rangle e^{-i E_{\nu} t} \langle\nu|n'l'm\rangle \right] \,.
\end{align}
Because of the block-diagonal structure in $m$, the full determinant factorizes into
\begin{align}
	S(t) = \prod_{m\geq 0} \left[ \det M^{(m)}_{(nl);(n'l')} \right]^{2-\delta_{m0}} \,,
\end{align}
where the factor $2-\delta_{m0}$ accounts for the degeneracy of $|m|$ and $-|m|$ sectors. Furthermore, since even and odd parity channels decouple, the determinant within each $m$ sector factorizes further into even and odd parity contributions, yielding the product structure
\begin{align}
	S(t) = S_\mathrm{even}(t) \cdot S_\mathrm{odd}(t) \,,
\end{align}
established in the main text. The impurity spectral function $A(\omega)$ is then obtained by Fourier transforming $S(t)$. All results are obtained in a spherical box of radius $R=20\, k^{-1}_F$ with $N_r=2000$ radial grid points and partial waves up to $l_\mathrm{max} = 6$, with a spectral broadening of $\eta=0.1\,E_F$.

\section{Comparison to the 1ph-Chevy ansatz} \label{app:chevy-ansatz}

As a complementary approach, we compute the polaron spectrum using the one particle-hole (1ph) Chevy ansatz~\cite{PhysRevA.74.063628}. Working in momentum space, the Hamiltonian for a static impurity takes the form
\begin{align} \label{eq:momentum-space-hamiltonian}
    \hat{H} = \sum_{\mathbf{k}} \varepsilon_{\mathbf{k}}\,
    \hat{c}^{\dagger}_{\mathbf{k}} \hat{c}_{\mathbf{k}}
    + \sum_{\mathbf{kk'}} V(\mathbf{k}-\mathbf{k}')\,
    \hat{c}^{\dagger}_{\mathbf{k}} \hat{c}_{\mathbf{k'}} \,,
\end{align}
with the free dispersion $\varepsilon_{\mathbf{k}} = \mathbf{k}^2/(2m)$ and the
dipolar impurity potential in momentum space
\begin{align} \label{eq:model-potential-momentum}
    V(\mathbf{k}-\mathbf{k}') = 4\pi d^2
    \left(3\cos^2\theta_{\mathbf{k}-\mathbf{k}'}-1\right)
    g(|\mathbf{k}-\mathbf{k}'|) \,,
\end{align}
where $g(q) = \big(8\sqrt{2}\,F(\tfrac{qR_0}{2\sqrt{2}})-8\,F(\tfrac{qR_0}{2})\big)/(q^3R^3_0)$ is expressed through the Dawson function $F(x)$, and $\theta_{\mathbf{k}-\mathbf{k}'}$ is the angle between $\mathbf{k}-\mathbf{k}'$ and the polarization axis. The variational ansatz for the polaron state is~\cite{PhysRevA.103.033324}
\begin{align} \label{eq:LLP-chevy-ansatz-SM}
    |\Psi\rangle = \alpha_0\, |\mathrm{FS}\rangle
    + \sum_{\mathbf{k}>,\mathbf{q}<} \alpha_{\mathbf{kq}}\,
    \hat{c}^{\dagger}_{\mathbf{k}}\hat{c}_{\mathbf{q}}|\mathrm{FS}\rangle \,,
\end{align}
where $\alpha_0$ and $\alpha_{\mathbf{kq}}$ are variational parameters, and $\mathbf{k}>$ ($\mathbf{q}<$) denotes momenta above (below) the Fermi surface.

Minimizing the energy $E = \langle\Psi|\hat{H}|\Psi\rangle / \langle\Psi|\Psi\rangle$ yields the coupled linear equations
\begin{align}
    E\,\alpha_0 &= \sum_{\mathbf{k}>,\mathbf{q}<}
    V(\mathbf{k}-\mathbf{q})\, \alpha_{\mathbf{k}\mathbf{q}} \,,
    \label{eq:chevy-gen-1}\\[1ex]
    E\,\alpha_{\mathbf{k}\mathbf{q}} &= \Omega_{\mathbf{k}\mathbf{q}}\,
    \alpha_{\mathbf{k}\mathbf{q}}
    + V(\mathbf{k}-\mathbf{q})\, \alpha_0
    + \sum_{\mathbf{k}'>} V(\mathbf{k}-\mathbf{k}')\,
    \alpha_{\mathbf{k}'\mathbf{q}} - \sum_{\mathbf{q}'<} V(\mathbf{q}-\mathbf{q}')\,
    \alpha_{\mathbf{k}\mathbf{q}'} \,,
    \label{eq:chevy-gen-2}
\end{align}
where $\Omega_{\mathbf{k}\mathbf{q}} = \varepsilon_{\mathbf{k}} - \varepsilon_{\mathbf{q}}$ is the particle-hole excitation energy. The three interaction terms describe, respectively, the coupling of a particle-hole pair back to the unperturbed Fermi sea, the scattering of the excited particle ($\mathbf{k}'\!\to\!\mathbf{k}$), and the scattering of the hole ($\mathbf{q}'\!\to\!\mathbf{q}$). The diagonal Hartree contributions $\propto V(\mathbf{0})$ vanish identically, since the angular average of $(3\cos^2\theta-1)$ is zero for the pure dipolar potential.

To exploit the rotational symmetry about the polarization axis, we expand the particle-hole amplitude in partial waves,
\begin{equation} \label{eq:pw-expansion}
    \alpha_{\mathbf{k}\mathbf{q}} = \sum_{m\in\mathbb{Z}}
    \sum_{l,l'\in\mathbb{N}_0} e^{im\phi}\,P^m_l(y)\,P^m_{l'}(y')\,
    \tilde{\alpha}_{kqmll'} \,,
\end{equation}
where $y=\cos\theta_{\mathbf{k}}$ and $y'=\cos\theta_{\mathbf{q}}$ are the polar angles of the particle and hole momenta, $\phi$ is their relative azimuthal angle, and $P^m_l$ are the associated Legendre polynomials. Because the total azimuthal angular momentum is conserved at total momentum $P=0$, a single magnetic quantum number $m$ labels both the particle and the hole.

Since the potential is invariant under $m\to-m$, we introduce the symmetric and antisymmetric combinations~\cite{Tiene:2024}
\begin{align}
    \tilde{\alpha}_{kqmll'}^{(\pm)} = \frac{1}{2}
    \big(\tilde{\alpha}_{kqmll'} \pm \tilde{\alpha}_{kq(-m)ll'}\big) \,,
\end{align}
with $\tilde{\alpha}_{kq0ll'}^{(+)} = \tilde{\alpha}_{kq0ll'}$ and $\tilde{\alpha}_{kq0ll'}^{(-)} = 0$. Only the symmetric sector couples to $\alpha_0$, and it suffices to restrict to $m\geq 0$. Discretizing the radial momenta on a grid with measure $dV=dk\,dq$, the eigenvalue equations become
\begin{align}
    E\,\alpha_0 &= \frac{1}{(2\pi)^2}\sum_{k,q}\sum_{m\geq 0}\sum_{ll'}
    kq\,\sqrt{dV}\,\mathcal{W}^m_{ll'}\, \tilde{V}_{kqmll'}\,
    \tilde{\alpha}_{kqmll'}^{(+)} \,, \label{eq:chevy-final-1}\\[1ex]
    E\,\tilde{\alpha}_{kqmll'}^{(+)} &= \Omega_{kq}\,
    \tilde{\alpha}_{kqmll'}^{(+)}
    + \frac{kq\,\sqrt{dV}}{(2\pi)^2}\,\mathcal{W}^m_{ll'}\,
    \tilde{V}_{kqmll'}\, \alpha_0 + \frac{k\sqrt{dk\,\mathcal{N}^m_l}}{(2\pi)^2}
    \sum_{k',\tilde{l}} k'\sqrt{dk'\,\mathcal{N}^m_{\tilde{l}}}\,
    \tilde{V}_{kk'ml\tilde{l}}\,
    \tilde{\alpha}_{k'qm\tilde{l}l'}^{(+)} \notag \\
    &\quad - \frac{q\sqrt{dq\,\mathcal{N}^m_{l'}}}{(2\pi)^2}
    \sum_{q',\tilde{l}'} q'\sqrt{dq'\,\mathcal{N}^m_{\tilde{l}'}}\,
    \tilde{V}_{qq'ml'\tilde{l}'}\,
    \tilde{\alpha}_{kq'ml\tilde{l}'}^{(+)} \,, \label{eq:chevy-final-2}
\end{align}
where $\mathcal{N}^m_l = \tfrac{2}{2l+1}\tfrac{(l+m)!}{(l-m)!}$ is the norm of the associated Legendre polynomials, $\mathcal{W}^m_{ll'} = \sqrt{(2-\delta_{m0})\,\mathcal{N}^m_l\mathcal{N}^m_{l'}}$ collects the degeneracy and normalization factors, and the $\sqrt{dV}$, $\sqrt{dk}$ factors render the matrix symmetric.

The angular-momentum matrix elements of the interaction are real-valued and evaluated numerically,
\begin{align} \label{eq:V-matrix-element}
    \tilde{V}_{kk'mll'} = \int_0^{2\pi} \!\frac{d\phi\,d\phi'}{(2\pi)^2}\,
    \frac{e^{im(\phi-\phi')}}{\mathcal{N}^m_l\mathcal{N}^m_{l'}}
    \int_{-1}^{1} \!dy\,dy'\, P^m_l(y) P^m_{l'}(y')\,
    V(\mathbf{k}-\mathbf{k}') \,.
\end{align}

The spectral function is computed via the Haydock recursion method~\cite{Haydock1972,PhysRevB.105.125404}, which avoids the explicit diagonalization of the full matrix and is particularly efficient for the large momentum-space matrices arising from the anisotropic dipolar potential. We discretize the particle (hole) momenta on a grid of $N_k=13$ ($N_q=11$) points up to a cutoff $k_\mathrm{max} = 40/R_0$, and retain partial waves up to $l_\mathrm{max} = 6$ and magnetic quantum numbers up to $m_\mathrm{max}=2$ with a broadening of $\eta = 0.1\,E_F$.

In the Chevy ansatz, both even and odd particle-hole excitations couple to the same amplitude $\alpha_0$ in Eq.~(\ref{eq:chevy-final-1}). This common coupling entangles the two parity sectors: rather than allowing the even- and odd-wave bound states to be occupied independently, the 1ph ansatz produces a single wavefunction in which they appear in superposition. Since capturing the simultaneous occupation of both bound states requires at least two particle-hole excitations, this is an intrinsic limitation of the 1ph truncation for static impurities in the dipolar setting, see Fig.~\ref{fig:6} in the main text.

\section{Universal behavior of $a_{02}$ near the $s$-wave resonance} \label{app:dipolar-dip}

Here, we derive the universal behavior of $a_{02}$ [see Eq.~\eqref{eq:a02}] and the associated cross section $\sigma_{02}$ near an $s$-wave resonance. We consider only the $l=0$ and $l=2$ channels and apply a distorted-wave Born approximation for the long-range part, treating the $s$-wave scattering length $a$ as a non-perturbative short-range parameter. The $K$-matrix in the $\{l=0,\, l=2\}$ subspace takes the form
\begin{equation}
    K = \begin{pmatrix} -k a & c k a_d \\ c k a_d &0 \end{pmatrix} \,,
\end{equation}
where the off-diagonal element arises from the Born approximation for the $l=0 \leftrightarrow l=2$ coupling, with the coefficient
\begin{equation}
    c = \frac{1}{3\sqrt{5}} \,,
\end{equation}
obtained from the angular projections onto the partial waves (see also first term of Eq.~(6) in Ref.~\cite{Fabrikant:1984} with $Q=2a_d$). The $S$-matrix is given by $S=(I+iK) (I-iK)^{-1}$, which results in
\begin{equation}
   S= \frac{1}{1+ika+c^2 k^2 a_d^2} \begin{pmatrix} 1-ika-c^2 k^2 a_d^2 & 2 c i k a_d \\ 2 c i k a_d &1+ ika -c^2k^2 a_d^2 \end{pmatrix} \,.
\end{equation}
For the $S_{02}$ component we thus get
\begin{equation}
    S_{02}=\frac{2 c i k a_d}{1+ika+c^2 k^2 a_d^2} = \frac{2 c i k a_d (1-ika +c^2 k^2 a_d^2)}{(1+c^2 k^2 a_d^2)^2+k^2 a^2} \,.
\end{equation}
So we find
\begin{align}
    \text{Re}\,S_{02}=\frac{2c k^2 a_d a}{(1+c^2 k^2 a_d^2)^2+k^2 a^2} \,, \quad
    \text{Im}\,S_{02}=\frac{2c k a_d(1 + c^2 k^2 a_d^2)}{(1+c^2 k^2 a_d^2)^2+k^2 a^2} \,.
\end{align}
Since the off-diagonal $K$-matrix element is evaluated only to linear order in $k a_d \ll 1$, it is more consistent to retain only terms to this order, giving
\begin{align}
    \text{Re}\,S_{02}\approx\frac{2c k^2 a_d a}{1+k^2 a^2} \,, \qquad
    \text{Im}\,S_{02}\approx\frac{2c k a_d}{1+k^2 a^2} \,. 
\end{align}
Since $T_{02}=iS_{02}$, the dipolar scattering length $a_{02}=T_{02}/(2k)$ follows as
\begin{equation}
    \text{Re}\,a_{02} = -\frac{\text{Im}\,S_{02}}{2k} \approx \frac{a_{02}^{\mathrm{Born}}}{1+k^2a^2} \,, \qquad
    \text{Im}\,a_{02} = \frac{\text{Re}\,S_{02}}{2k} \approx \frac{-k a\, a_{02}^{\mathrm{Born}}}{1+k^2a^2} \,,
\end{equation}
with $a_{02}^{\mathrm{Born}}=-c a_d=-a_d/(3\sqrt{5})$, reproducing Eqs.~\eqref{eq:Born-length} and~\eqref{eq:dipolar-length}. The characteristic dip in $\text{Re}\,a_{02}$ at the $s$-wave resonance results from destructive interference between the direct dipolar coupling and the contribution dressed by resonant $s$-wave scattering. Indeed, to first order in $a_d$, $S_{02}\propto 1+S_{00}$, which vanishes in the unitarity limit $S_{00}\to-1$ ($\delta_{00}\to\pm\pi/2$).  The width of the dip is set by $k|a|$ and therefore vanishes as $k\to 0$, consistent with the numerical results shown in Fig.~\ref{fig:2}(b).

To compute the collision cross section, we need
\begin{equation}
    \sigma = \frac{\pi |1-S|^2}{k^2} \,.
\end{equation}
For the value of $\sigma_{02}$, we therefore require the value of $|S_{02}|^2$. We find
\begin{equation}
    |S_{02}|^2 \approx \frac{4 c^2 k^4 a^2_d a^2 +4 c^2 k^2 a_d^2}{(1+k^2 a^2)^2}=\frac{4 c^2 k^2 a^2_d}{1+k^2 a^2}=4 c^2 k^2 a^2_d |\cos(\tan^{-1}(ka))|^2 = 4 c^2 k^2 a^2_d |\cos(\delta_{00})|^2 \,.
\end{equation}
This universal form is relevant for thermalization dynamics in dipolar gases, where the cross section governs the rate of momentum transfer between different angular-momentum components.

\end{widetext}
%

%


\begin{thebibliography}{79}%
\makeatletter
\providecommand \@ifxundefined [1]{%
 \@ifx{#1\undefined}
}%
\providecommand \@ifnum [1]{%
 \ifnum #1\expandafter \@firstoftwo
 \else \expandafter \@secondoftwo
 \fi
}%
\providecommand \@ifx [1]{%
 \ifx #1\expandafter \@firstoftwo
 \else \expandafter \@secondoftwo
 \fi
}%
\providecommand \natexlab [1]{#1}%
\providecommand \enquote  [1]{``#1''}%
\providecommand \bibnamefont  [1]{#1}%
\providecommand \bibfnamefont [1]{#1}%
\providecommand \citenamefont [1]{#1}%
\providecommand \href@noop [0]{\@secondoftwo}%
\providecommand \href [0]{\begingroup \@sanitize@url \@href}%
\providecommand \@href[1]{\@@startlink{#1}\@@href}%
\providecommand \@@href[1]{\endgroup#1\@@endlink}%
\providecommand \@sanitize@url [0]{\catcode `\\12\catcode `\$12\catcode
  `\&12\catcode `\#12\catcode `\^12\catcode `\_12\catcode `\%12\relax}%
\providecommand \@@startlink[1]{}%
\providecommand \@@endlink[0]{}%
\providecommand \url  [0]{\begingroup\@sanitize@url \@url }%
\providecommand \@url [1]{\endgroup\@href {#1}{\urlprefix }}%
\providecommand \urlprefix  [0]{URL }%
\providecommand \Eprint [0]{\href }%
\providecommand \doibase [0]{https://doi.org/}%
\providecommand \selectlanguage [0]{\@gobble}%
\providecommand \bibinfo  [0]{\@secondoftwo}%
\providecommand \bibfield  [0]{\@secondoftwo}%
\providecommand \translation [1]{[#1]}%
\providecommand \BibitemOpen [0]{}%
\providecommand \bibitemStop [0]{}%
\providecommand \bibitemNoStop [0]{.\EOS\space}%
\providecommand \EOS [0]{\spacefactor3000\relax}%
\providecommand \BibitemShut  [1]{\csname bibitem#1\endcsname}%
\let\auto@bib@innerbib\@empty
\bibitem [{\citenamefont {Schmidt}\ \emph
  {et~al.}(2018{\natexlab{a}})\citenamefont {Schmidt}, \citenamefont {Knap},
  \citenamefont {Ivanov}, \citenamefont {You}, \citenamefont {Cetina},\ and\
  \citenamefont {Demler}}]{Schmidt2018}%
  \BibitemOpen
  \bibfield  {author} {\bibinfo {author} {\bibfnamefont {R.}~\bibnamefont
  {Schmidt}}, \bibinfo {author} {\bibfnamefont {M.}~\bibnamefont {Knap}},
  \bibinfo {author} {\bibfnamefont {D.~A.}\ \bibnamefont {Ivanov}}, \bibinfo
  {author} {\bibfnamefont {J.-S.}\ \bibnamefont {You}}, \bibinfo {author}
  {\bibfnamefont {M.}~\bibnamefont {Cetina}},\ and\ \bibinfo {author}
  {\bibfnamefont {E.}~\bibnamefont {Demler}},\ }\bibfield  {title} {\bibinfo
  {title} {{Universal many-body response of heavy impurities coupled to a Fermi
  sea: a review of recent progress}},\ }\href
  {https://doi.org/10.1088/1361-6633/aa9593} {\bibfield  {journal} {\bibinfo
  {journal} {Rep. Prog. Phys.}\ }\textbf {\bibinfo {volume} {81}},\ \bibinfo
  {pages} {024401} (\bibinfo {year} {2018}{\natexlab{a}})}\BibitemShut
  {NoStop}%
\bibitem [{\citenamefont {Massignan}\ \emph {et~al.}(2026)\citenamefont
  {Massignan}, \citenamefont {Schmidt}, \citenamefont {Astrakharchik},
  \citenamefont {{\.I}mamoglu}, \citenamefont {Zwierlein}, \citenamefont
  {Arlt},\ and\ \citenamefont {Bruun}}]{Massignan:2025uee}%
  \BibitemOpen
  \bibfield  {author} {\bibinfo {author} {\bibfnamefont {P.}~\bibnamefont
  {Massignan}}, \bibinfo {author} {\bibfnamefont {R.}~\bibnamefont {Schmidt}},
  \bibinfo {author} {\bibfnamefont {G.~E.}\ \bibnamefont {Astrakharchik}},
  \bibinfo {author} {\bibfnamefont {A.}~\bibnamefont {{\.I}mamoglu}}, \bibinfo
  {author} {\bibfnamefont {M.}~\bibnamefont {Zwierlein}}, \bibinfo {author}
  {\bibfnamefont {J.~J.}\ \bibnamefont {Arlt}},\ and\ \bibinfo {author}
  {\bibfnamefont {G.~M.}\ \bibnamefont {Bruun}},\ }\bibfield  {title} {\bibinfo
  {title} {{Polarons in atomic gases and two-dimensional semiconductors}},\
  }\href {https://doi.org/10.1103/4nng-bb9z} {\bibfield  {journal} {\bibinfo
  {journal} {Rev. Mod. Phys.}\ }\textbf {\bibinfo {volume} {98}},\ \bibinfo
  {pages} {035002} (\bibinfo {year} {2026})}\BibitemShut {NoStop}%
\bibitem [{\citenamefont {Massignan}\ \emph {et~al.}(2014)\citenamefont
  {Massignan}, \citenamefont {Zaccanti},\ and\ \citenamefont
  {Bruun}}]{Massignan:2014}%
  \BibitemOpen
  \bibfield  {author} {\bibinfo {author} {\bibfnamefont {P.}~\bibnamefont
  {Massignan}}, \bibinfo {author} {\bibfnamefont {M.}~\bibnamefont
  {Zaccanti}},\ and\ \bibinfo {author} {\bibfnamefont {G.~M.}\ \bibnamefont
  {Bruun}},\ }\bibfield  {title} {\bibinfo {title} {{Polarons, dressed
  molecules and itinerant ferromagnetism in ultracold Fermi gases}},\ }\href
  {https://doi.org/10.1088/0034-4885/77/3/034401} {\bibfield  {journal}
  {\bibinfo  {journal} {Rep. Prog. Phys.}\ }\textbf {\bibinfo {volume} {77}},\
  \bibinfo {pages} {034401} (\bibinfo {year} {2014})}\BibitemShut {NoStop}%
\bibitem [{\citenamefont {Scazza}\ \emph {et~al.}(2022)\citenamefont {Scazza},
  \citenamefont {Zaccanti}, \citenamefont {Massignan}, \citenamefont {Parish},\
  and\ \citenamefont {Levinsen}}]{Scazza:2022bez}%
  \BibitemOpen
  \bibfield  {author} {\bibinfo {author} {\bibfnamefont {F.}~\bibnamefont
  {Scazza}}, \bibinfo {author} {\bibfnamefont {M.}~\bibnamefont {Zaccanti}},
  \bibinfo {author} {\bibfnamefont {P.}~\bibnamefont {Massignan}}, \bibinfo
  {author} {\bibfnamefont {M.~M.}\ \bibnamefont {Parish}},\ and\ \bibinfo
  {author} {\bibfnamefont {J.}~\bibnamefont {Levinsen}},\ }\bibfield  {title}
  {\bibinfo {title} {{Repulsive Fermi and Bose Polarons in Quantum Gases}},\
  }\href {https://doi.org/10.3390/atoms10020055} {\bibfield  {journal}
  {\bibinfo  {journal} {Atoms}\ }\textbf {\bibinfo {volume} {10}},\ \bibinfo
  {pages} {55} (\bibinfo {year} {2022})}\BibitemShut {NoStop}%
\bibitem [{\citenamefont {Schirotzek}\ \emph {et~al.}(2009)\citenamefont
  {Schirotzek}, \citenamefont {Wu}, \citenamefont {Sommer},\ and\ \citenamefont
  {Zwierlein}}]{Schirotzek:2009exp}%
  \BibitemOpen
  \bibfield  {author} {\bibinfo {author} {\bibfnamefont {A.}~\bibnamefont
  {Schirotzek}}, \bibinfo {author} {\bibfnamefont {C.-H.}\ \bibnamefont {Wu}},
  \bibinfo {author} {\bibfnamefont {A.}~\bibnamefont {Sommer}},\ and\ \bibinfo
  {author} {\bibfnamefont {M.~W.}\ \bibnamefont {Zwierlein}},\ }\bibfield
  {title} {\bibinfo {title} {{Observation of Fermi Polarons in a Tunable Fermi
  Liquid of Ultracold Atoms}},\ }\href
  {https://doi.org/10.1103/PhysRevLett.102.230402} {\bibfield  {journal}
  {\bibinfo  {journal} {Phys. Rev. Lett.}\ }\textbf {\bibinfo {volume} {102}},\
  \bibinfo {pages} {230402} (\bibinfo {year} {2009})}\BibitemShut {NoStop}%
\bibitem [{\citenamefont {Punk}\ \emph {et~al.}(2009)\citenamefont {Punk},
  \citenamefont {Dumitrescu},\ and\ \citenamefont {Zwerger}}]{Punk:2009}%
  \BibitemOpen
  \bibfield  {author} {\bibinfo {author} {\bibfnamefont {M.}~\bibnamefont
  {Punk}}, \bibinfo {author} {\bibfnamefont {P.~T.}\ \bibnamefont
  {Dumitrescu}},\ and\ \bibinfo {author} {\bibfnamefont {W.}~\bibnamefont
  {Zwerger}},\ }\bibfield  {title} {\bibinfo {title} {{Polaron-to-molecule
  transition in a strongly imbalanced Fermi gas}},\ }\href
  {https://doi.org/10.1103/PhysRevA.80.053605} {\bibfield  {journal} {\bibinfo
  {journal} {Phys. Rev. A}\ }\textbf {\bibinfo {volume} {80}},\ \bibinfo
  {pages} {053605} (\bibinfo {year} {2009})}\BibitemShut {NoStop}%
\bibitem [{\citenamefont {Chevy}\ and\ \citenamefont
  {Mora}(2010)}]{ChevyMora2010}%
  \BibitemOpen
  \bibfield  {author} {\bibinfo {author} {\bibfnamefont {F.}~\bibnamefont
  {Chevy}}\ and\ \bibinfo {author} {\bibfnamefont {C.}~\bibnamefont {Mora}},\
  }\bibfield  {title} {\bibinfo {title} {{Ultra-cold polarized Fermi gases}},\
  }\href {https://doi.org/10.1088/0034-4885/73/11/112401} {\bibfield  {journal}
  {\bibinfo  {journal} {Rep. Prog. Phys.}\ }\textbf {\bibinfo {volume} {73}},\
  \bibinfo {pages} {112401} (\bibinfo {year} {2010})}\BibitemShut {NoStop}%
\bibitem [{\citenamefont {Chomaz}\ \emph {et~al.}(2023)\citenamefont {Chomaz},
  \citenamefont {Ferrier-Barbut}, \citenamefont {Ferlaino}, \citenamefont
  {Laburthe-Tolra}, \citenamefont {Lev},\ and\ \citenamefont
  {Pfau}}]{Chomaz:2023}%
  \BibitemOpen
  \bibfield  {author} {\bibinfo {author} {\bibfnamefont {L.}~\bibnamefont
  {Chomaz}}, \bibinfo {author} {\bibfnamefont {I.}~\bibnamefont
  {Ferrier-Barbut}}, \bibinfo {author} {\bibfnamefont {F.}~\bibnamefont
  {Ferlaino}}, \bibinfo {author} {\bibfnamefont {B.}~\bibnamefont
  {Laburthe-Tolra}}, \bibinfo {author} {\bibfnamefont {B.~L.}\ \bibnamefont
  {Lev}},\ and\ \bibinfo {author} {\bibfnamefont {T.}~\bibnamefont {Pfau}},\
  }\bibfield  {title} {\bibinfo {title} {{Dipolar physics: a review of
  experiments with magnetic quantum gases}},\ }\href
  {https://doi.org/10.1088/1361-6633/aca814} {\bibfield  {journal} {\bibinfo
  {journal} {Rep. Prog. Phys.}\ }\textbf {\bibinfo {volume} {86}},\ \bibinfo
  {pages} {026401} (\bibinfo {year} {2023})}\BibitemShut {NoStop}%
\bibitem [{\citenamefont {De~Marco}\ \emph {et~al.}(2019)\citenamefont
  {De~Marco}, \citenamefont {Valtolina}, \citenamefont {Matsuda}, \citenamefont
  {Tobias}, \citenamefont {Covey},\ and\ \citenamefont {Ye}}]{demarco:2019}%
  \BibitemOpen
  \bibfield  {author} {\bibinfo {author} {\bibfnamefont {L.}~\bibnamefont
  {De~Marco}}, \bibinfo {author} {\bibfnamefont {G.}~\bibnamefont {Valtolina}},
  \bibinfo {author} {\bibfnamefont {K.}~\bibnamefont {Matsuda}}, \bibinfo
  {author} {\bibfnamefont {W.~G.}\ \bibnamefont {Tobias}}, \bibinfo {author}
  {\bibfnamefont {J.~P.}\ \bibnamefont {Covey}},\ and\ \bibinfo {author}
  {\bibfnamefont {J.}~\bibnamefont {Ye}},\ }\bibfield  {title} {\bibinfo
  {title} {{A degenerate Fermi gas of polar molecules}},\ }\href
  {https://doi.org/10.1126/science.aau7230} {\bibfield  {journal} {\bibinfo
  {journal} {Science}\ }\textbf {\bibinfo {volume} {363}},\ \bibinfo {pages}
  {853} (\bibinfo {year} {2019})}\BibitemShut {NoStop}%
\bibitem [{\citenamefont {Schindewolf}\ \emph {et~al.}(2022)\citenamefont
  {Schindewolf}, \citenamefont {Bause}, \citenamefont {Chen}, \citenamefont
  {Duda}, \citenamefont {Karman}, \citenamefont {Bloch},\ and\ \citenamefont
  {Luo}}]{schindewolf:2022}%
  \BibitemOpen
  \bibfield  {author} {\bibinfo {author} {\bibfnamefont {A.}~\bibnamefont
  {Schindewolf}}, \bibinfo {author} {\bibfnamefont {R.}~\bibnamefont {Bause}},
  \bibinfo {author} {\bibfnamefont {X.-Y.}\ \bibnamefont {Chen}}, \bibinfo
  {author} {\bibfnamefont {M.}~\bibnamefont {Duda}}, \bibinfo {author}
  {\bibfnamefont {T.}~\bibnamefont {Karman}}, \bibinfo {author} {\bibfnamefont
  {I.}~\bibnamefont {Bloch}},\ and\ \bibinfo {author} {\bibfnamefont {X.-Y.}\
  \bibnamefont {Luo}},\ }\bibfield  {title} {\bibinfo {title} {{Evaporation of
  microwave-shielded polar molecules to quantum degeneracy}},\ }\href
  {https://doi.org/10.1038/s41586-022-04914-8} {\bibfield  {journal} {\bibinfo
  {journal} {Nature}\ }\textbf {\bibinfo {volume} {607}},\ \bibinfo {pages}
  {677} (\bibinfo {year} {2022})}\BibitemShut {NoStop}%
\bibitem [{\citenamefont {Bigagli}\ \emph {et~al.}(2024)\citenamefont
  {Bigagli}, \citenamefont {Yuan}, \citenamefont {Zhang}, \citenamefont
  {Bulatovic}, \citenamefont {Karman}, \citenamefont {Stevenson},\ and\
  \citenamefont {Will}}]{Bigagli:2024}%
  \BibitemOpen
  \bibfield  {author} {\bibinfo {author} {\bibfnamefont {N.}~\bibnamefont
  {Bigagli}}, \bibinfo {author} {\bibfnamefont {W.}~\bibnamefont {Yuan}},
  \bibinfo {author} {\bibfnamefont {S.}~\bibnamefont {Zhang}}, \bibinfo
  {author} {\bibfnamefont {B.}~\bibnamefont {Bulatovic}}, \bibinfo {author}
  {\bibfnamefont {T.}~\bibnamefont {Karman}}, \bibinfo {author} {\bibfnamefont
  {I.}~\bibnamefont {Stevenson}},\ and\ \bibinfo {author} {\bibfnamefont
  {S.}~\bibnamefont {Will}},\ }\bibfield  {title} {\bibinfo {title}
  {{Observation of Bose–Einstein condensation of dipolar molecules}},\ }\href
  {https://doi.org/10.1038/s41586-024-07492-z} {\bibfield  {journal} {\bibinfo
  {journal} {Nature}\ }\textbf {\bibinfo {volume} {631}},\ \bibinfo {pages}
  {289} (\bibinfo {year} {2024})}\BibitemShut {NoStop}%
\bibitem [{\citenamefont {Schindewolf}\ \emph {et~al.}(2026)\citenamefont
  {Schindewolf}, \citenamefont {Hertkorn}, \citenamefont {Stevenson},
  \citenamefont {Ciardi}, \citenamefont {Gro{\ss{}}}, \citenamefont {Wang},
  \citenamefont {Karman}, \citenamefont {Qu{\'e}m{\'e}ner}, \citenamefont
  {Will}, \citenamefont {Pohl},\ and\ \citenamefont
  {Langen}}]{schindewolf:2026}%
  \BibitemOpen
  \bibfield  {author} {\bibinfo {author} {\bibfnamefont {A.}~\bibnamefont
  {Schindewolf}}, \bibinfo {author} {\bibfnamefont {J.}~\bibnamefont
  {Hertkorn}}, \bibinfo {author} {\bibfnamefont {I.}~\bibnamefont {Stevenson}},
  \bibinfo {author} {\bibfnamefont {M.}~\bibnamefont {Ciardi}}, \bibinfo
  {author} {\bibfnamefont {P.}~\bibnamefont {Gro{\ss{}}}}, \bibinfo {author}
  {\bibfnamefont {D.}~\bibnamefont {Wang}}, \bibinfo {author} {\bibfnamefont
  {T.}~\bibnamefont {Karman}}, \bibinfo {author} {\bibfnamefont
  {G.}~\bibnamefont {Qu{\'e}m{\'e}ner}}, \bibinfo {author} {\bibfnamefont
  {S.}~\bibnamefont {Will}}, \bibinfo {author} {\bibfnamefont {T.}~\bibnamefont
  {Pohl}},\ and\ \bibinfo {author} {\bibfnamefont {T.}~\bibnamefont {Langen}},\
  }\bibfield  {title} {\bibinfo {title} {{Colloquium: Strongly dipolar
  molecular Bose-Einstein condensates: From few- to many-body physics}},\
  }\href {https://doi.org/10.1103/r55l-f93m} {\bibfield  {journal} {\bibinfo
  {journal} {Rev. Mod. Phys.}\ }\textbf {\bibinfo {volume} {98}},\ \bibinfo
  {pages} {031002} (\bibinfo {year} {2026})}\BibitemShut {NoStop}%
\bibitem [{\citenamefont {Bohn}\ \emph {et~al.}(2009)\citenamefont {Bohn},
  \citenamefont {Cavagnero},\ and\ \citenamefont {Ticknor}}]{Bohn:2009}%
  \BibitemOpen
  \bibfield  {author} {\bibinfo {author} {\bibfnamefont {J.~L.}\ \bibnamefont
  {Bohn}}, \bibinfo {author} {\bibfnamefont {M.}~\bibnamefont {Cavagnero}},\
  and\ \bibinfo {author} {\bibfnamefont {C.}~\bibnamefont {Ticknor}},\
  }\bibfield  {title} {\bibinfo {title} {{Quasi-universal dipolar scattering in
  cold and ultracold gases}},\ }\href
  {https://doi.org/10.1088/1367-2630/11/5/055039} {\bibfield  {journal}
  {\bibinfo  {journal} {New J. Phys.}\ }\textbf {\bibinfo {volume} {11}},\
  \bibinfo {pages} {055039} (\bibinfo {year} {2009})}\BibitemShut {NoStop}%
\bibitem [{\citenamefont {Bohn}\ and\ \citenamefont {Jin}(2014)}]{Bohn:2014}%
  \BibitemOpen
  \bibfield  {author} {\bibinfo {author} {\bibfnamefont {J.~L.}\ \bibnamefont
  {Bohn}}\ and\ \bibinfo {author} {\bibfnamefont {D.~S.}\ \bibnamefont {Jin}},\
  }\bibfield  {title} {\bibinfo {title} {{Differential scattering and
  rethermalization in ultracold dipolar gases}},\ }\href
  {https://doi.org/10.1103/PhysRevA.89.022702} {\bibfield  {journal} {\bibinfo
  {journal} {Phys. Rev. A}\ }\textbf {\bibinfo {volume} {89}},\ \bibinfo
  {pages} {022702} (\bibinfo {year} {2014})}\BibitemShut {NoStop}%
\bibitem [{\citenamefont {Lu}\ \emph {et~al.}(2012)\citenamefont {Lu},
  \citenamefont {Burdick},\ and\ \citenamefont {Lev}}]{PhysRevLett.108.215301}%
  \BibitemOpen
  \bibfield  {author} {\bibinfo {author} {\bibfnamefont {M.}~\bibnamefont
  {Lu}}, \bibinfo {author} {\bibfnamefont {N.~Q.}\ \bibnamefont {Burdick}},\
  and\ \bibinfo {author} {\bibfnamefont {B.~L.}\ \bibnamefont {Lev}},\
  }\bibfield  {title} {\bibinfo {title} {{Quantum Degenerate Dipolar Fermi
  Gas}},\ }\href {https://doi.org/10.1103/PhysRevLett.108.215301} {\bibfield
  {journal} {\bibinfo  {journal} {Phys. Rev. Lett.}\ }\textbf {\bibinfo
  {volume} {108}},\ \bibinfo {pages} {215301} (\bibinfo {year}
  {2012})}\BibitemShut {NoStop}%
\bibitem [{\citenamefont {Baranov}\ \emph {et~al.}(2012)\citenamefont
  {Baranov}, \citenamefont {Dalmonte}, \citenamefont {Pupillo},\ and\
  \citenamefont {Zoller}}]{Baranov:2012}%
  \BibitemOpen
  \bibfield  {author} {\bibinfo {author} {\bibfnamefont {M.~A.}\ \bibnamefont
  {Baranov}}, \bibinfo {author} {\bibfnamefont {M.}~\bibnamefont {Dalmonte}},
  \bibinfo {author} {\bibfnamefont {G.}~\bibnamefont {Pupillo}},\ and\ \bibinfo
  {author} {\bibfnamefont {P.}~\bibnamefont {Zoller}},\ }\bibfield  {title}
  {\bibinfo {title} {{Condensed Matter Theory of Dipolar Quantum Gases}},\
  }\href {https://doi.org/10.1021/cr2003568} {\bibfield  {journal} {\bibinfo
  {journal} {Chem. Rev.}\ }\textbf {\bibinfo {volume} {112}},\ \bibinfo {pages}
  {5012} (\bibinfo {year} {2012})}\BibitemShut {NoStop}%
\bibitem [{\citenamefont {Lahaye}\ \emph {et~al.}(2009)\citenamefont {Lahaye},
  \citenamefont {Menotti}, \citenamefont {Santos}, \citenamefont {Lewenstein},\
  and\ \citenamefont {Pfau}}]{Lahaye:2009qqr}%
  \BibitemOpen
  \bibfield  {author} {\bibinfo {author} {\bibfnamefont {T.}~\bibnamefont
  {Lahaye}}, \bibinfo {author} {\bibfnamefont {C.}~\bibnamefont {Menotti}},
  \bibinfo {author} {\bibfnamefont {L.}~\bibnamefont {Santos}}, \bibinfo
  {author} {\bibfnamefont {M.}~\bibnamefont {Lewenstein}},\ and\ \bibinfo
  {author} {\bibfnamefont {T.}~\bibnamefont {Pfau}},\ }\bibfield  {title}
  {\bibinfo {title} {{The physics of dipolar bosonic quantum gases}},\ }\href
  {https://doi.org/10.1088/0034-4885/72/12/126401} {\bibfield  {journal}
  {\bibinfo  {journal} {Rep. Prog. Phys.}\ }\textbf {\bibinfo {volume} {72}},\
  \bibinfo {pages} {126401} (\bibinfo {year} {2009})}\BibitemShut {NoStop}%
\bibitem [{\citenamefont {Baranov}\ \emph {et~al.}(2002)\citenamefont
  {Baranov}, \citenamefont {Mar'enko}, \citenamefont {Rychkov},\ and\
  \citenamefont {Shlyapnikov}}]{Baranov:2002}%
  \BibitemOpen
  \bibfield  {author} {\bibinfo {author} {\bibfnamefont {M.~A.}\ \bibnamefont
  {Baranov}}, \bibinfo {author} {\bibfnamefont {M.~S.}\ \bibnamefont
  {Mar'enko}}, \bibinfo {author} {\bibfnamefont {V.~S.}\ \bibnamefont
  {Rychkov}},\ and\ \bibinfo {author} {\bibfnamefont {G.~V.}\ \bibnamefont
  {Shlyapnikov}},\ }\bibfield  {title} {\bibinfo {title} {{Superfluid pairing
  in a polarized dipolar Fermi gas}},\ }\href
  {https://doi.org/10.1103/PhysRevA.66.013606} {\bibfield  {journal} {\bibinfo
  {journal} {Phys. Rev. A}\ }\textbf {\bibinfo {volume} {66}},\ \bibinfo
  {pages} {013606} (\bibinfo {year} {2002})}\BibitemShut {NoStop}%
\bibitem [{\citenamefont {Cooper}\ and\ \citenamefont
  {Shlyapnikov}(2009)}]{cooper:2009}%
  \BibitemOpen
  \bibfield  {author} {\bibinfo {author} {\bibfnamefont {N.~R.}\ \bibnamefont
  {Cooper}}\ and\ \bibinfo {author} {\bibfnamefont {G.~V.}\ \bibnamefont
  {Shlyapnikov}},\ }\bibfield  {title} {\bibinfo {title} {{Stable Topological
  Superfluid Phase of Ultracold Polar Fermionic Molecules}},\ }\href
  {https://doi.org/10.1103/PhysRevLett.103.155302} {\bibfield  {journal}
  {\bibinfo  {journal} {Phys. Rev. Lett.}\ }\textbf {\bibinfo {volume} {103}},\
  \bibinfo {pages} {155302} (\bibinfo {year} {2009})}\BibitemShut {NoStop}%
\bibitem [{\citenamefont {Levinsen}\ \emph {et~al.}(2011)\citenamefont
  {Levinsen}, \citenamefont {Cooper},\ and\ \citenamefont
  {Shlyapnikov}}]{levinsen:2011}%
  \BibitemOpen
  \bibfield  {author} {\bibinfo {author} {\bibfnamefont {J.}~\bibnamefont
  {Levinsen}}, \bibinfo {author} {\bibfnamefont {N.~R.}\ \bibnamefont
  {Cooper}},\ and\ \bibinfo {author} {\bibfnamefont {G.~V.}\ \bibnamefont
  {Shlyapnikov}},\ }\bibfield  {title} {\bibinfo {title} {{Topological $p_x +
  ip_y$ superfluid phase of fermionic polar molecules}},\ }\href
  {https://doi.org/10.1103/PhysRevA.84.013603} {\bibfield  {journal} {\bibinfo
  {journal} {Phys. Rev. A}\ }\textbf {\bibinfo {volume} {84}},\ \bibinfo
  {pages} {013603} (\bibinfo {year} {2011})}\BibitemShut {NoStop}%
\bibitem [{\citenamefont {Deng}\ \emph {et~al.}(2023)\citenamefont {Deng},
  \citenamefont {Chen}, \citenamefont {Luo}, \citenamefont {Zhang},
  \citenamefont {Yi},\ and\ \citenamefont {Shi}}]{deng:2023}%
  \BibitemOpen
  \bibfield  {author} {\bibinfo {author} {\bibfnamefont {F.}~\bibnamefont
  {Deng}}, \bibinfo {author} {\bibfnamefont {X.-Y.}\ \bibnamefont {Chen}},
  \bibinfo {author} {\bibfnamefont {X.-Y.}\ \bibnamefont {Luo}}, \bibinfo
  {author} {\bibfnamefont {W.}~\bibnamefont {Zhang}}, \bibinfo {author}
  {\bibfnamefont {S.}~\bibnamefont {Yi}},\ and\ \bibinfo {author}
  {\bibfnamefont {T.}~\bibnamefont {Shi}},\ }\bibfield  {title} {\bibinfo
  {title} {{Effective Potential and Superfluidity of Microwave-Shielded Polar
  Molecules}},\ }\href {https://doi.org/10.1103/PhysRevLett.130.183001}
  {\bibfield  {journal} {\bibinfo  {journal} {Phys. Rev. Lett.}\ }\textbf
  {\bibinfo {volume} {130}},\ \bibinfo {pages} {183001} (\bibinfo {year}
  {2023})}\BibitemShut {NoStop}%
\bibitem [{\citenamefont {Christensen}\ \emph {et~al.}(2022)\citenamefont
  {Christensen}, \citenamefont {Camacho-Guardian},\ and\ \citenamefont
  {Bruun}}]{PhysRevA.105.023309}%
  \BibitemOpen
  \bibfield  {author} {\bibinfo {author} {\bibfnamefont {E.~R.}\ \bibnamefont
  {Christensen}}, \bibinfo {author} {\bibfnamefont {A.}~\bibnamefont
  {Camacho-Guardian}},\ and\ \bibinfo {author} {\bibfnamefont {G.~M.}\
  \bibnamefont {Bruun}},\ }\bibfield  {title} {\bibinfo {title} {{Mobile ion in
  a Fermi sea}},\ }\href {https://doi.org/10.1103/PhysRevA.105.023309}
  {\bibfield  {journal} {\bibinfo  {journal} {Phys. Rev. A}\ }\textbf {\bibinfo
  {volume} {105}},\ \bibinfo {pages} {023309} (\bibinfo {year}
  {2022})}\BibitemShut {NoStop}%
\bibitem [{\citenamefont {Camargo}\ \emph {et~al.}(2018)\citenamefont
  {Camargo}, \citenamefont {Schmidt}, \citenamefont {Whalen}, \citenamefont
  {Ding}, \citenamefont {Woehl}, \citenamefont {Yoshida}, \citenamefont
  {Burgd{\"o}rfer}, \citenamefont {Dunning}, \citenamefont {Sadeghpour},
  \citenamefont {Demler},\ and\ \citenamefont {Pfau}}]{Camargo:2018}%
  \BibitemOpen
  \bibfield  {author} {\bibinfo {author} {\bibfnamefont {F.}~\bibnamefont
  {Camargo}}, \bibinfo {author} {\bibfnamefont {R.}~\bibnamefont {Schmidt}},
  \bibinfo {author} {\bibfnamefont {J.~D.}\ \bibnamefont {Whalen}}, \bibinfo
  {author} {\bibfnamefont {R.}~\bibnamefont {Ding}}, \bibinfo {author}
  {\bibfnamefont {G.}~\bibnamefont {Woehl}}, \bibinfo {author} {\bibfnamefont
  {S.}~\bibnamefont {Yoshida}}, \bibinfo {author} {\bibfnamefont
  {J.}~\bibnamefont {Burgd{\"o}rfer}}, \bibinfo {author} {\bibfnamefont
  {F.~B.}\ \bibnamefont {Dunning}}, \bibinfo {author} {\bibfnamefont {H.~R.}\
  \bibnamefont {Sadeghpour}}, \bibinfo {author} {\bibfnamefont
  {E.}~\bibnamefont {Demler}},\ and\ \bibinfo {author} {\bibfnamefont
  {T.}~\bibnamefont {Pfau}},\ }\bibfield  {title} {\bibinfo {title} {{Creation
  of Rydberg Polarons in a Bose Gas}},\ }\href
  {https://doi.org/10.1103/PhysRevLett.120.083401} {\bibfield  {journal}
  {\bibinfo  {journal} {Phys. Rev. Lett.}\ }\textbf {\bibinfo {volume} {120}},\
  \bibinfo {pages} {083401} (\bibinfo {year} {2018})}\BibitemShut {NoStop}%
\bibitem [{\citenamefont {Schmidt}\ \emph
  {et~al.}(2018{\natexlab{b}})\citenamefont {Schmidt}, \citenamefont
  {Sadeghpour},\ and\ \citenamefont {Demler}}]{Schmidt:2018rydberg}%
  \BibitemOpen
  \bibfield  {author} {\bibinfo {author} {\bibfnamefont {R.}~\bibnamefont
  {Schmidt}}, \bibinfo {author} {\bibfnamefont {H.~R.}\ \bibnamefont
  {Sadeghpour}},\ and\ \bibinfo {author} {\bibfnamefont {E.}~\bibnamefont
  {Demler}},\ }\bibfield  {title} {\bibinfo {title} {{Theory of excitation of
  Rydberg polarons in an atomic quantum gas}},\ }\href
  {https://doi.org/10.1103/PhysRevA.97.022707} {\bibfield  {journal} {\bibinfo
  {journal} {Phys. Rev. A}\ }\textbf {\bibinfo {volume} {97}},\ \bibinfo
  {pages} {022707} (\bibinfo {year} {2018}{\natexlab{b}})}\BibitemShut
  {NoStop}%
\bibitem [{\citenamefont {Whalen}\ \emph {et~al.}(2019)\citenamefont {Whalen},
  \citenamefont {Kanungo}, \citenamefont {Ding}, \citenamefont {Wagner},
  \citenamefont {Schmidt}, \citenamefont {Yoshida}, \citenamefont
  {Burgd{\"o}rfer}, \citenamefont {Becker}, \citenamefont {Kleinbach},
  \citenamefont {Pfau},\ and\ \citenamefont {Dunning}}]{Whalen:2019}%
  \BibitemOpen
  \bibfield  {author} {\bibinfo {author} {\bibfnamefont {J.~D.}\ \bibnamefont
  {Whalen}}, \bibinfo {author} {\bibfnamefont {S.~K.}\ \bibnamefont {Kanungo}},
  \bibinfo {author} {\bibfnamefont {R.}~\bibnamefont {Ding}}, \bibinfo {author}
  {\bibfnamefont {M.}~\bibnamefont {Wagner}}, \bibinfo {author} {\bibfnamefont
  {R.}~\bibnamefont {Schmidt}}, \bibinfo {author} {\bibfnamefont
  {S.}~\bibnamefont {Yoshida}}, \bibinfo {author} {\bibfnamefont
  {J.}~\bibnamefont {Burgd{\"o}rfer}}, \bibinfo {author} {\bibfnamefont
  {S.}~\bibnamefont {Becker}}, \bibinfo {author} {\bibfnamefont {K.~S.}\
  \bibnamefont {Kleinbach}}, \bibinfo {author} {\bibfnamefont {T.}~\bibnamefont
  {Pfau}},\ and\ \bibinfo {author} {\bibfnamefont {F.~B.}\ \bibnamefont
  {Dunning}},\ }\bibfield  {title} {\bibinfo {title} {{Probing nonlocal spatial
  correlations in quantum gases with ultra-long-range Rydberg molecules}},\
  }\href {https://doi.org/10.1103/PhysRevA.100.011402} {\bibfield  {journal}
  {\bibinfo  {journal} {Phys. Rev. A}\ }\textbf {\bibinfo {volume} {100}},\
  \bibinfo {pages} {011402} (\bibinfo {year} {2019})}\BibitemShut {NoStop}%
\bibitem [{\citenamefont {Sous}\ \emph {et~al.}(2020)\citenamefont {Sous},
  \citenamefont {Sadeghpour}, \citenamefont {Killian}, \citenamefont {Demler},\
  and\ \citenamefont {Schmidt}}]{Sous2020}%
  \BibitemOpen
  \bibfield  {author} {\bibinfo {author} {\bibfnamefont {J.}~\bibnamefont
  {Sous}}, \bibinfo {author} {\bibfnamefont {H.~R.}\ \bibnamefont
  {Sadeghpour}}, \bibinfo {author} {\bibfnamefont {T.~C.}\ \bibnamefont
  {Killian}}, \bibinfo {author} {\bibfnamefont {E.}~\bibnamefont {Demler}},\
  and\ \bibinfo {author} {\bibfnamefont {R.}~\bibnamefont {Schmidt}},\
  }\bibfield  {title} {\bibinfo {title} {{Rydberg impurity in a Fermi gas:
  Quantum statistics and rotational blockade}},\ }\href
  {https://doi.org/10.1103/PhysRevResearch.2.023021} {\bibfield  {journal}
  {\bibinfo  {journal} {Phys. Rev. Res.}\ }\textbf {\bibinfo {volume} {2}},\
  \bibinfo {pages} {023021} (\bibinfo {year} {2020})}\BibitemShut {NoStop}%
\bibitem [{\citenamefont {Gievers}\ \emph {et~al.}(2024)\citenamefont
  {Gievers}, \citenamefont {Wagner},\ and\ \citenamefont
  {Schmidt}}]{Gievers2024}%
  \BibitemOpen
  \bibfield  {author} {\bibinfo {author} {\bibfnamefont {M.}~\bibnamefont
  {Gievers}}, \bibinfo {author} {\bibfnamefont {M.}~\bibnamefont {Wagner}},\
  and\ \bibinfo {author} {\bibfnamefont {R.}~\bibnamefont {Schmidt}},\
  }\bibfield  {title} {\bibinfo {title} {{Probing Polaron Clouds by Rydberg
  Atom Spectroscopy}},\ }\href {https://doi.org/10.1103/PhysRevLett.132.053401}
  {\bibfield  {journal} {\bibinfo  {journal} {Phys. Rev. Lett.}\ }\textbf
  {\bibinfo {volume} {132}},\ \bibinfo {pages} {053401} (\bibinfo {year}
  {2024})}\BibitemShut {NoStop}%
\bibitem [{\citenamefont {Durst}\ and\ \citenamefont
  {Eiles}(2024)}]{Durst:2024ryd}%
  \BibitemOpen
  \bibfield  {author} {\bibinfo {author} {\bibfnamefont {A.~A.~T.}\
  \bibnamefont {Durst}}\ and\ \bibinfo {author} {\bibfnamefont {M.~T.}\
  \bibnamefont {Eiles}},\ }\bibfield  {title} {\bibinfo {title} {{Phenomenology
  of a Rydberg impurity in an ideal Bose-Einstein condensate}},\ }\href
  {https://doi.org/10.1103/PhysRevResearch.6.L042009} {\bibfield  {journal}
  {\bibinfo  {journal} {Phys. Rev. Res.}\ }\textbf {\bibinfo {volume} {6}},\
  \bibinfo {pages} {L042009} (\bibinfo {year} {2024})}\BibitemShut {NoStop}%
\bibitem [{\citenamefont {Durst}\ \emph {et~al.}(2026)\citenamefont {Durst},
  \citenamefont {Rittenhouse}, \citenamefont {Sadeghpour},\ and\ \citenamefont
  {Eiles}}]{Durst:2026anisotropic}%
  \BibitemOpen
  \bibfield  {author} {\bibinfo {author} {\bibfnamefont {A.~A.~T.}\
  \bibnamefont {Durst}}, \bibinfo {author} {\bibfnamefont {S.~T.}\ \bibnamefont
  {Rittenhouse}}, \bibinfo {author} {\bibfnamefont {H.~R.}\ \bibnamefont
  {Sadeghpour}},\ and\ \bibinfo {author} {\bibfnamefont {M.~T.}\ \bibnamefont
  {Eiles}},\ }\bibfield  {title} {\bibinfo {title} {{Anisotropic and
  Nonadditive Interactions of a Rydberg Impurity in a Quantum Bath}},\ }\href
  {https://doi.org/10.1103/38vj-6fqc} {\bibfield  {journal} {\bibinfo
  {journal} {Phys. Rev. Lett.}\ }\textbf {\bibinfo {volume} {137}},\ \bibinfo
  {pages} {073401} (\bibinfo {year} {2026})}\BibitemShut {NoStop}%
\bibitem [{\citenamefont {Shi}\ \emph {et~al.}(2012)\citenamefont {Shi},
  \citenamefont {Qi},\ and\ \citenamefont {Zhai}}]{PhysRevA.85.020702}%
  \BibitemOpen
  \bibfield  {author} {\bibinfo {author} {\bibfnamefont {Z.-Y.}\ \bibnamefont
  {Shi}}, \bibinfo {author} {\bibfnamefont {R.}~\bibnamefont {Qi}},\ and\
  \bibinfo {author} {\bibfnamefont {H.}~\bibnamefont {Zhai}},\ }\bibfield
  {title} {\bibinfo {title} {{$s$-wave-scattering resonances induced by dipolar
  interactions of polar molecules}},\ }\href
  {https://doi.org/10.1103/PhysRevA.85.020702} {\bibfield  {journal} {\bibinfo
  {journal} {Phys. Rev. A}\ }\textbf {\bibinfo {volume} {85}},\ \bibinfo
  {pages} {020702} (\bibinfo {year} {2012})}\BibitemShut {NoStop}%
\bibitem [{\citenamefont {Nishimura}\ \emph {et~al.}(2021)\citenamefont
  {Nishimura}, \citenamefont {Nakano}, \citenamefont {Iida}, \citenamefont
  {Tajima}, \citenamefont {Miyakawa},\ and\ \citenamefont
  {Yabu}}]{PhysRevA.103.033324}%
  \BibitemOpen
  \bibfield  {author} {\bibinfo {author} {\bibfnamefont {K.}~\bibnamefont
  {Nishimura}}, \bibinfo {author} {\bibfnamefont {E.}~\bibnamefont {Nakano}},
  \bibinfo {author} {\bibfnamefont {K.}~\bibnamefont {Iida}}, \bibinfo {author}
  {\bibfnamefont {H.}~\bibnamefont {Tajima}}, \bibinfo {author} {\bibfnamefont
  {T.}~\bibnamefont {Miyakawa}},\ and\ \bibinfo {author} {\bibfnamefont
  {H.}~\bibnamefont {Yabu}},\ }\bibfield  {title} {\bibinfo {title} {{Ground
  state of the polaron in an ultracold dipolar Fermi gas}},\ }\href
  {https://doi.org/10.1103/PhysRevA.103.033324} {\bibfield  {journal} {\bibinfo
   {journal} {Phys. Rev. A}\ }\textbf {\bibinfo {volume} {103}},\ \bibinfo
  {pages} {033324} (\bibinfo {year} {2021})}\BibitemShut {NoStop}%
\bibitem [{\citenamefont {Volosniev}\ \emph {et~al.}(2023)\citenamefont
  {Volosniev}, \citenamefont {Bighin}, \citenamefont {Santos},\ and\
  \citenamefont {Pe\~na Ardila}}]{Volosniev:2023xfn}%
  \BibitemOpen
  \bibfield  {author} {\bibinfo {author} {\bibfnamefont {A.~G.}\ \bibnamefont
  {Volosniev}}, \bibinfo {author} {\bibfnamefont {G.}~\bibnamefont {Bighin}},
  \bibinfo {author} {\bibfnamefont {L.}~\bibnamefont {Santos}},\ and\ \bibinfo
  {author} {\bibfnamefont {L.~A.}\ \bibnamefont {Pe\~na Ardila}},\ }\bibfield
  {title} {\bibinfo {title} {{Non-equilibrium dynamics of dipolar polarons}},\
  }\href {https://doi.org/10.21468/SciPostPhys.15.6.232} {\bibfield  {journal}
  {\bibinfo  {journal} {SciPost Phys.}\ }\textbf {\bibinfo {volume} {15}},\
  \bibinfo {pages} {232} (\bibinfo {year} {2023})}\BibitemShut {NoStop}%
\bibitem [{\citenamefont {Nakano}\ \emph {et~al.}(2024)\citenamefont {Nakano},
  \citenamefont {Miyakawa},\ and\ \citenamefont {Yabu}}]{Nakano:2024the}%
  \BibitemOpen
  \bibfield  {author} {\bibinfo {author} {\bibfnamefont {E.}~\bibnamefont
  {Nakano}}, \bibinfo {author} {\bibfnamefont {T.}~\bibnamefont {Miyakawa}},\
  and\ \bibinfo {author} {\bibfnamefont {H.}~\bibnamefont {Yabu}},\ }\bibfield
  {title} {\bibinfo {title} {{Two-body problem of impurity atoms in dipolar
  Fermi gas}},\ }\href@noop {} {\  (\bibinfo {year} {2024})},\ \Eprint
  {https://arxiv.org/abs/2404.14866} {arXiv:2404.14866 [cond-mat.quant-gas]}
  \BibitemShut {NoStop}%
\bibitem [{\citenamefont {Tiene}\ \emph {et~al.}(2024)\citenamefont {Tiene},
  \citenamefont {Bracho}, \citenamefont {Parish}, \citenamefont {Levinsen},\
  and\ \citenamefont {Marchetti}}]{Tiene:2024}%
  \BibitemOpen
  \bibfield  {author} {\bibinfo {author} {\bibfnamefont {A.}~\bibnamefont
  {Tiene}}, \bibinfo {author} {\bibfnamefont {A.~T.}\ \bibnamefont {Bracho}},
  \bibinfo {author} {\bibfnamefont {M.~M.}\ \bibnamefont {Parish}}, \bibinfo
  {author} {\bibfnamefont {J.}~\bibnamefont {Levinsen}},\ and\ \bibinfo
  {author} {\bibfnamefont {F.~M.}\ \bibnamefont {Marchetti}},\ }\bibfield
  {title} {\bibinfo {title} {{Multiple polaron quasiparticles with dipolar
  fermions in a bilayer geometry}},\ }\href
  {https://doi.org/10.1103/PhysRevA.109.033318} {\bibfield  {journal} {\bibinfo
   {journal} {Phys. Rev. A}\ }\textbf {\bibinfo {volume} {109}},\ \bibinfo
  {pages} {033318} (\bibinfo {year} {2024})}\BibitemShut {NoStop}%
\bibitem [{\citenamefont {Kain}\ and\ \citenamefont
  {Ling}(2014)}]{PhysRevA.89.023612}%
  \BibitemOpen
  \bibfield  {author} {\bibinfo {author} {\bibfnamefont {B.}~\bibnamefont
  {Kain}}\ and\ \bibinfo {author} {\bibfnamefont {H.~Y.}\ \bibnamefont
  {Ling}},\ }\bibfield  {title} {\bibinfo {title} {{Polarons in a dipolar
  condensate}},\ }\href {https://doi.org/10.1103/PhysRevA.89.023612} {\bibfield
   {journal} {\bibinfo  {journal} {Phys. Rev. A}\ }\textbf {\bibinfo {volume}
  {89}},\ \bibinfo {pages} {023612} (\bibinfo {year} {2014})}\BibitemShut
  {NoStop}%
\bibitem [{\citenamefont {Shukla}\ \emph {et~al.}(2024)\citenamefont {Shukla},
  \citenamefont {Volosniev},\ and\ \citenamefont {Armstrong}}]{Shukla2024}%
  \BibitemOpen
  \bibfield  {author} {\bibinfo {author} {\bibfnamefont {N.}~\bibnamefont
  {Shukla}}, \bibinfo {author} {\bibfnamefont {A.~G.}\ \bibnamefont
  {Volosniev}},\ and\ \bibinfo {author} {\bibfnamefont {J.~R.}\ \bibnamefont
  {Armstrong}},\ }\bibfield  {title} {\bibinfo {title} {{Anisotropic potential
  immersed in a dipolar Bose-Einstein condensate}},\ }\href
  {https://doi.org/10.1103/PhysRevA.110.053317} {\bibfield  {journal} {\bibinfo
   {journal} {Phys. Rev. A}\ }\textbf {\bibinfo {volume} {110}},\ \bibinfo
  {pages} {053317} (\bibinfo {year} {2024})}\BibitemShut {NoStop}%
\bibitem [{\citenamefont {Shukla}\ and\ \citenamefont
  {Armstrong}(2025)}]{Shukla:2025}%
  \BibitemOpen
  \bibfield  {author} {\bibinfo {author} {\bibfnamefont {N.}~\bibnamefont
  {Shukla}}\ and\ \bibinfo {author} {\bibfnamefont {J.~R.}\ \bibnamefont
  {Armstrong}},\ }\bibfield  {title} {\bibinfo {title} {{Properties of a Static
  Dipolar Impurity in a 2D Dipolar BEC}},\ }\href
  {https://doi.org/10.3390/atoms13030024} {\bibfield  {journal} {\bibinfo
  {journal} {Atoms}\ }\textbf {\bibinfo {volume} {13}},\ \bibinfo {pages} {24}
  (\bibinfo {year} {2025})}\BibitemShut {NoStop}%
\bibitem [{\citenamefont {Anderson}(1967)}]{Anderson:1967}%
  \BibitemOpen
  \bibfield  {author} {\bibinfo {author} {\bibfnamefont {P.~W.}\ \bibnamefont
  {Anderson}},\ }\bibfield  {title} {\bibinfo {title} {{Infrared Catastrophe in
  Fermi Gases with Local Scattering Potentials}},\ }\href
  {https://doi.org/10.1103/PhysRevLett.18.1049} {\bibfield  {journal} {\bibinfo
   {journal} {Phys. Rev. Lett.}\ }\textbf {\bibinfo {volume} {18}},\ \bibinfo
  {pages} {1049} (\bibinfo {year} {1967})}\BibitemShut {NoStop}%
\bibitem [{\citenamefont {Fumi}(1955)}]{Fumi:1955}%
  \BibitemOpen
  \bibfield  {author} {\bibinfo {author} {\bibfnamefont {F.~G.}\ \bibnamefont
  {Fumi}},\ }\bibfield  {title} {\bibinfo {title} {{CXVI. Vacancies in
  monovalent metals}},\ }\href {https://doi.org/10.1080/14786440908520622}
  {\bibfield  {journal} {\bibinfo  {journal} {Philos. Mag.}\ }\textbf {\bibinfo
  {volume} {46}},\ \bibinfo {pages} {1007} (\bibinfo {year}
  {1955})}\BibitemShut {NoStop}%
\bibitem [{\citenamefont {Mahan}(2000)}]{Mahan2000}%
  \BibitemOpen
  \bibfield  {author} {\bibinfo {author} {\bibfnamefont {G.~D.}\ \bibnamefont
  {Mahan}},\ }\href {https://doi.org/10.1007/978-1-4757-5714-9} {\emph
  {\bibinfo {title} {{Many-Particle Physics}}}},\ \bibinfo {edition} {3rd}\
  ed.\ (\bibinfo  {publisher} {Springer New York, NY},\ \bibinfo {year}
  {2000})\BibitemShut {NoStop}%
\bibitem [{\citenamefont {Xiang}\ and\ \citenamefont {Wu}(2022)}]{Xiang2022}%
  \BibitemOpen
  \bibfield  {author} {\bibinfo {author} {\bibfnamefont {T.}~\bibnamefont
  {Xiang}}\ and\ \bibinfo {author} {\bibfnamefont {C.}~\bibnamefont {Wu}},\
  }\href {https://doi.org/10.1017/9781009218566} {\emph {\bibinfo {title}
  {{D-wave Superconductivity}}}}\ (\bibinfo  {publisher} {Cambridge University
  Press},\ \bibinfo {year} {2022})\BibitemShut {NoStop}%
\bibitem [{\citenamefont {Ticknor}(2009)}]{PhysRevA.80.052702}%
  \BibitemOpen
  \bibfield  {author} {\bibinfo {author} {\bibfnamefont {C.}~\bibnamefont
  {Ticknor}},\ }\bibfield  {title} {\bibinfo {title} {{Two-dimensional dipolar
  scattering}},\ }\href {https://doi.org/10.1103/PhysRevA.80.052702} {\bibfield
   {journal} {\bibinfo  {journal} {Phys. Rev. A}\ }\textbf {\bibinfo {volume}
  {80}},\ \bibinfo {pages} {052702} (\bibinfo {year} {2009})}\BibitemShut
  {NoStop}%
\bibitem [{\citenamefont {Ticknor}(2010)}]{PhysRevA.81.042708}%
  \BibitemOpen
  \bibfield  {author} {\bibinfo {author} {\bibfnamefont {C.}~\bibnamefont
  {Ticknor}},\ }\bibfield  {title} {\bibinfo {title} {{Quasi-two-dimensional
  dipolar scattering}},\ }\href {https://doi.org/10.1103/PhysRevA.81.042708}
  {\bibfield  {journal} {\bibinfo  {journal} {Phys. Rev. A}\ }\textbf {\bibinfo
  {volume} {81}},\ \bibinfo {pages} {042708} (\bibinfo {year}
  {2010})}\BibitemShut {NoStop}%
\bibitem [{\citenamefont {Cherny}(2023)}]{PhysRevA.107.033315}%
  \BibitemOpen
  \bibfield  {author} {\bibinfo {author} {\bibfnamefont {A.~Y.}\ \bibnamefont
  {Cherny}},\ }\bibfield  {title} {\bibinfo {title} {{Adiabatic sweep theorem
  for three-dimensional dipolar Bose gases}},\ }\href
  {https://doi.org/10.1103/PhysRevA.107.033315} {\bibfield  {journal} {\bibinfo
   {journal} {Phys. Rev. A}\ }\textbf {\bibinfo {volume} {107}},\ \bibinfo
  {pages} {033315} (\bibinfo {year} {2023})}\BibitemShut {NoStop}%
\bibitem [{\citenamefont {Crubellier}\ \emph {et~al.}(2019)\citenamefont
  {Crubellier}, \citenamefont {Gonz\'alez-F\'erez}, \citenamefont {Koch},\ and\
  \citenamefont {Luc-Koenig}}]{PhysRevA.99.032709}%
  \BibitemOpen
  \bibfield  {author} {\bibinfo {author} {\bibfnamefont {A.}~\bibnamefont
  {Crubellier}}, \bibinfo {author} {\bibfnamefont {R.}~\bibnamefont
  {Gonz\'alez-F\'erez}}, \bibinfo {author} {\bibfnamefont {C.~P.}\ \bibnamefont
  {Koch}},\ and\ \bibinfo {author} {\bibfnamefont {E.}~\bibnamefont
  {Luc-Koenig}},\ }\bibfield  {title} {\bibinfo {title} {{Defining the $p$-wave
  scattering volume in the presence of dipolar interactions}},\ }\href
  {https://doi.org/10.1103/PhysRevA.99.032709} {\bibfield  {journal} {\bibinfo
  {journal} {Phys. Rev. A}\ }\textbf {\bibinfo {volume} {99}},\ \bibinfo
  {pages} {032709} (\bibinfo {year} {2019})}\BibitemShut {NoStop}%
\bibitem [{\citenamefont {Karman}\ \emph {et~al.}(2025)\citenamefont {Karman},
  \citenamefont {Bigagli}, \citenamefont {Yuan}, \citenamefont {Zhang},
  \citenamefont {Stevenson},\ and\ \citenamefont {Will}}]{Karman2025DMS}%
  \BibitemOpen
  \bibfield  {author} {\bibinfo {author} {\bibfnamefont {T.}~\bibnamefont
  {Karman}}, \bibinfo {author} {\bibfnamefont {N.}~\bibnamefont {Bigagli}},
  \bibinfo {author} {\bibfnamefont {W.}~\bibnamefont {Yuan}}, \bibinfo {author}
  {\bibfnamefont {S.}~\bibnamefont {Zhang}}, \bibinfo {author} {\bibfnamefont
  {I.}~\bibnamefont {Stevenson}},\ and\ \bibinfo {author} {\bibfnamefont
  {S.}~\bibnamefont {Will}},\ }\bibfield  {title} {\bibinfo {title} {{Double
  Microwave Shielding}},\ }\href {https://doi.org/10.1103/b8pm-3prn} {\bibfield
   {journal} {\bibinfo  {journal} {PRX Quantum}\ }\textbf {\bibinfo {volume}
  {6}},\ \bibinfo {pages} {020358} (\bibinfo {year} {2025})}\BibitemShut
  {NoStop}%
\bibitem [{\citenamefont {Chevy}(2006)}]{PhysRevA.74.063628}%
  \BibitemOpen
  \bibfield  {author} {\bibinfo {author} {\bibfnamefont {F.}~\bibnamefont
  {Chevy}},\ }\bibfield  {title} {\bibinfo {title} {{Universal phase diagram of
  a strongly interacting Fermi gas with unbalanced spin populations}},\ }\href
  {https://doi.org/10.1103/PhysRevA.74.063628} {\bibfield  {journal} {\bibinfo
  {journal} {Phys. Rev. A}\ }\textbf {\bibinfo {volume} {74}},\ \bibinfo
  {pages} {063628} (\bibinfo {year} {2006})}\BibitemShut {NoStop}%
\bibitem [{\citenamefont {Qi}\ \emph {et~al.}(2013)\citenamefont {Qi},
  \citenamefont {Shi},\ and\ \citenamefont {Zhai}}]{PhysRevLett.110.045302}%
  \BibitemOpen
  \bibfield  {author} {\bibinfo {author} {\bibfnamefont {R.}~\bibnamefont
  {Qi}}, \bibinfo {author} {\bibfnamefont {Z.-Y.}\ \bibnamefont {Shi}},\ and\
  \bibinfo {author} {\bibfnamefont {H.}~\bibnamefont {Zhai}},\ }\bibfield
  {title} {\bibinfo {title} {{Fermion Pairing Across a Dipolar Interaction
  Induced Resonance}},\ }\href {https://doi.org/10.1103/PhysRevLett.110.045302}
  {\bibfield  {journal} {\bibinfo  {journal} {Phys. Rev. Lett.}\ }\textbf
  {\bibinfo {volume} {110}},\ \bibinfo {pages} {045302} (\bibinfo {year}
  {2013})}\BibitemShut {NoStop}%
\bibitem [{\citenamefont {Roudnev}\ and\ \citenamefont
  {Cavagnero}(2009)}]{Roudnev:2009}%
  \BibitemOpen
  \bibfield  {author} {\bibinfo {author} {\bibfnamefont {V.}~\bibnamefont
  {Roudnev}}\ and\ \bibinfo {author} {\bibfnamefont {M.}~\bibnamefont
  {Cavagnero}},\ }\bibfield  {title} {\bibinfo {title} {{Resonance phenomena in
  ultracold dipole–dipole scattering}},\ }\href
  {https://doi.org/10.1088/0953-4075/42/4/044017} {\bibfield  {journal}
  {\bibinfo  {journal} {J. Phys. B: At. Mol. Opt. Phys.}\ }\textbf {\bibinfo
  {volume} {42}},\ \bibinfo {pages} {044017} (\bibinfo {year}
  {2009})}\BibitemShut {NoStop}%
\bibitem [{\citenamefont {Wang}\ and\ \citenamefont {Greene}(2012)}]{Wang2012}%
  \BibitemOpen
  \bibfield  {author} {\bibinfo {author} {\bibfnamefont {Y.}~\bibnamefont
  {Wang}}\ and\ \bibinfo {author} {\bibfnamefont {C.~H.}\ \bibnamefont
  {Greene}},\ }\bibfield  {title} {\bibinfo {title} {{Universal bound and
  scattering properties for two dipoles}},\ }\href
  {https://doi.org/10.1103/PhysRevA.85.022704} {\bibfield  {journal} {\bibinfo
  {journal} {Phys. Rev. A}\ }\textbf {\bibinfo {volume} {85}},\ \bibinfo
  {pages} {022704} (\bibinfo {year} {2012})}\BibitemShut {NoStop}%
\bibitem [{\citenamefont {Karman}\ and\ \citenamefont
  {Hutson}(2018)}]{PhysRevLett.121.163401}%
  \BibitemOpen
  \bibfield  {author} {\bibinfo {author} {\bibfnamefont {T.}~\bibnamefont
  {Karman}}\ and\ \bibinfo {author} {\bibfnamefont {J.~M.}\ \bibnamefont
  {Hutson}},\ }\bibfield  {title} {\bibinfo {title} {{Microwave Shielding of
  Ultracold Polar Molecules}},\ }\href
  {https://doi.org/10.1103/PhysRevLett.121.163401} {\bibfield  {journal}
  {\bibinfo  {journal} {Phys. Rev. Lett.}\ }\textbf {\bibinfo {volume} {121}},\
  \bibinfo {pages} {163401} (\bibinfo {year} {2018})}\BibitemShut {NoStop}%
\bibitem [{\citenamefont {Chen}\ \emph {et~al.}(2023)\citenamefont {Chen},
  \citenamefont {Schindewolf}, \citenamefont {Eppelt}, \citenamefont {Bause},
  \citenamefont {Duda}, \citenamefont {Biswas}, \citenamefont {Karman},
  \citenamefont {Hilker}, \citenamefont {Bloch},\ and\ \citenamefont
  {Luo}}]{Chen:2022svd}%
  \BibitemOpen
  \bibfield  {author} {\bibinfo {author} {\bibfnamefont {X.-Y.}\ \bibnamefont
  {Chen}}, \bibinfo {author} {\bibfnamefont {A.}~\bibnamefont {Schindewolf}},
  \bibinfo {author} {\bibfnamefont {S.}~\bibnamefont {Eppelt}}, \bibinfo
  {author} {\bibfnamefont {R.}~\bibnamefont {Bause}}, \bibinfo {author}
  {\bibfnamefont {M.}~\bibnamefont {Duda}}, \bibinfo {author} {\bibfnamefont
  {S.}~\bibnamefont {Biswas}}, \bibinfo {author} {\bibfnamefont
  {T.}~\bibnamefont {Karman}}, \bibinfo {author} {\bibfnamefont
  {T.}~\bibnamefont {Hilker}}, \bibinfo {author} {\bibfnamefont
  {I.}~\bibnamefont {Bloch}},\ and\ \bibinfo {author} {\bibfnamefont {X.-Y.}\
  \bibnamefont {Luo}},\ }\bibfield  {title} {\bibinfo {title} {{Field-linked
  resonances of polar molecules}},\ }\href
  {https://doi.org/10.1038/s41586-022-05651-8} {\bibfield  {journal} {\bibinfo
  {journal} {Nature}\ }\textbf {\bibinfo {volume} {614}},\ \bibinfo {pages}
  {59} (\bibinfo {year} {2023})}\BibitemShut {NoStop}%
\bibitem [{\citenamefont {Chen}\ \emph {et~al.}(2024)\citenamefont {Chen},
  \citenamefont {Biswas}, \citenamefont {Eppelt}, \citenamefont {Schindewolf},
  \citenamefont {Deng}, \citenamefont {Shi}, \citenamefont {Yi}, \citenamefont
  {Hilker}, \citenamefont {Bloch},\ and\ \citenamefont {Luo}}]{Chen:2023dqs}%
  \BibitemOpen
  \bibfield  {author} {\bibinfo {author} {\bibfnamefont {X.-Y.}\ \bibnamefont
  {Chen}}, \bibinfo {author} {\bibfnamefont {S.}~\bibnamefont {Biswas}},
  \bibinfo {author} {\bibfnamefont {S.}~\bibnamefont {Eppelt}}, \bibinfo
  {author} {\bibfnamefont {A.}~\bibnamefont {Schindewolf}}, \bibinfo {author}
  {\bibfnamefont {F.}~\bibnamefont {Deng}}, \bibinfo {author} {\bibfnamefont
  {T.}~\bibnamefont {Shi}}, \bibinfo {author} {\bibfnamefont {S.}~\bibnamefont
  {Yi}}, \bibinfo {author} {\bibfnamefont {T.~A.}\ \bibnamefont {Hilker}},
  \bibinfo {author} {\bibfnamefont {I.}~\bibnamefont {Bloch}},\ and\ \bibinfo
  {author} {\bibfnamefont {X.-Y.}\ \bibnamefont {Luo}},\ }\bibfield  {title}
  {\bibinfo {title} {{Ultracold field-linked tetratomic molecules}},\ }\href
  {https://doi.org/10.1038/s41586-023-06986-6} {\bibfield  {journal} {\bibinfo
  {journal} {Nature}\ }\textbf {\bibinfo {volume} {626}},\ \bibinfo {pages}
  {283} (\bibinfo {year} {2024})}\BibitemShut {NoStop}%
\bibitem [{\citenamefont {Deng}\ \emph {et~al.}(2025)\citenamefont {Deng},
  \citenamefont {Hu}, \citenamefont {Jin}, \citenamefont {Yi},\ and\
  \citenamefont {Shi}}]{Deng:2025fjv}%
  \BibitemOpen
  \bibfield  {author} {\bibinfo {author} {\bibfnamefont {F.}~\bibnamefont
  {Deng}}, \bibinfo {author} {\bibfnamefont {X.}~\bibnamefont {Hu}}, \bibinfo
  {author} {\bibfnamefont {W.-J.}\ \bibnamefont {Jin}}, \bibinfo {author}
  {\bibfnamefont {S.}~\bibnamefont {Yi}},\ and\ \bibinfo {author}
  {\bibfnamefont {T.}~\bibnamefont {Shi}},\ }\bibfield  {title} {\bibinfo
  {title} {{Two- and many-body physics of ultracold molecules dressed by dual
  microwave fields}},\ }\href {https://doi.org/10.1038/s41467-025-66067-2}
  {\bibfield  {journal} {\bibinfo  {journal} {Nat. Commun.}\ }\textbf {\bibinfo
  {volume} {16}},\ \bibinfo {pages} {11219} (\bibinfo {year}
  {2025})}\BibitemShut {NoStop}%
\bibitem [{\citenamefont {O\l{}dziejewski}\ and\ \citenamefont
  {Jachymski}(2016)}]{oldziejweski:2016}%
  \BibitemOpen
  \bibfield  {author} {\bibinfo {author} {\bibfnamefont {R.}~\bibnamefont
  {O\l{}dziejewski}}\ and\ \bibinfo {author} {\bibfnamefont {K.}~\bibnamefont
  {Jachymski}},\ }\bibfield  {title} {\bibinfo {title} {{Properties of strongly
  dipolar Bose gases beyond the Born approximation}},\ }\href
  {https://doi.org/10.1103/PhysRevA.94.063638} {\bibfield  {journal} {\bibinfo
  {journal} {Phys. Rev. A}\ }\textbf {\bibinfo {volume} {94}},\ \bibinfo
  {pages} {063638} (\bibinfo {year} {2016})}\BibitemShut {NoStop}%
\bibitem [{\citenamefont {Kanjilal}\ \emph {et~al.}(2007)\citenamefont
  {Kanjilal}, \citenamefont {Bohn},\ and\ \citenamefont
  {Blume}}]{PhysRevA.75.052705}%
  \BibitemOpen
  \bibfield  {author} {\bibinfo {author} {\bibfnamefont {K.}~\bibnamefont
  {Kanjilal}}, \bibinfo {author} {\bibfnamefont {J.~L.}\ \bibnamefont {Bohn}},\
  and\ \bibinfo {author} {\bibfnamefont {D.}~\bibnamefont {Blume}},\ }\bibfield
   {title} {\bibinfo {title} {{Pseudopotential treatment of two aligned dipoles
  under external harmonic confinement}},\ }\href
  {https://doi.org/10.1103/PhysRevA.75.052705} {\bibfield  {journal} {\bibinfo
  {journal} {Phys. Rev. A}\ }\textbf {\bibinfo {volume} {75}},\ \bibinfo
  {pages} {052705} (\bibinfo {year} {2007})}\BibitemShut {NoStop}%
\bibitem [{\citenamefont {Karman}\ \emph {et~al.}(2018)\citenamefont {Karman},
  \citenamefont {Frye}, \citenamefont {Reddel},\ and\ \citenamefont
  {Hutson}}]{PhysRevA.98.062502}%
  \BibitemOpen
  \bibfield  {author} {\bibinfo {author} {\bibfnamefont {T.}~\bibnamefont
  {Karman}}, \bibinfo {author} {\bibfnamefont {M.~D.}\ \bibnamefont {Frye}},
  \bibinfo {author} {\bibfnamefont {J.~D.}\ \bibnamefont {Reddel}},\ and\
  \bibinfo {author} {\bibfnamefont {J.~M.}\ \bibnamefont {Hutson}},\ }\bibfield
   {title} {\bibinfo {title} {{Near-threshold bound states of the dipole-dipole
  interaction}},\ }\href {https://doi.org/10.1103/PhysRevA.98.062502}
  {\bibfield  {journal} {\bibinfo  {journal} {Phys. Rev. A}\ }\textbf {\bibinfo
  {volume} {98}},\ \bibinfo {pages} {062502} (\bibinfo {year}
  {2018})}\BibitemShut {NoStop}%
\bibitem [{\citenamefont {O'Malley}\ \emph {et~al.}(1961)\citenamefont
  {O'Malley}, \citenamefont {Spruch},\ and\ \citenamefont
  {Rosenberg}}]{omalley:1961}%
  \BibitemOpen
  \bibfield  {author} {\bibinfo {author} {\bibfnamefont {T.~F.}\ \bibnamefont
  {O'Malley}}, \bibinfo {author} {\bibfnamefont {L.}~\bibnamefont {Spruch}},\
  and\ \bibinfo {author} {\bibfnamefont {L.}~\bibnamefont {Rosenberg}},\
  }\bibfield  {title} {\bibinfo {title} {{Modification of Effective‐Range
  Theory in the Presence of a Long‐Range ($r^{-4}$) Potential}},\ }\href
  {https://doi.org/10.1063/1.1703735} {\bibfield  {journal} {\bibinfo
  {journal} {J. Math. Phys.}\ }\textbf {\bibinfo {volume} {2}},\ \bibinfo
  {pages} {491} (\bibinfo {year} {1961})}\BibitemShut {NoStop}%
\bibitem [{\citenamefont {Fabrikant}(1981)}]{Fabrikant:1981}%
  \BibitemOpen
  \bibfield  {author} {\bibinfo {author} {\bibfnamefont {I.~I.}\ \bibnamefont
  {Fabrikant}},\ }\bibfield  {title} {\bibinfo {title} {{Partial-wave analysis
  of low-energy electron scattering by quadrupole molecules}},\ }\href
  {https://doi.org/10.1088/0022-3700/14/2/016} {\bibfield  {journal} {\bibinfo
  {journal} {J. Phys. B: Atom. Mol. Phys.}\ }\textbf {\bibinfo {volume} {14}},\
  \bibinfo {pages} {335} (\bibinfo {year} {1981})}\BibitemShut {NoStop}%
\bibitem [{\citenamefont {Fabrikant}(1984)}]{Fabrikant:1984}%
  \BibitemOpen
  \bibfield  {author} {\bibinfo {author} {\bibfnamefont {I.~I.}\ \bibnamefont
  {Fabrikant}},\ }\bibfield  {title} {\bibinfo {title} {{Effective-range
  analysis of low-energy electron scattering by non-polar molecules}},\ }\href
  {https://doi.org/10.1088/0022-3700/17/20/020} {\bibfield  {journal} {\bibinfo
   {journal} {J. Phys. B: Atom. Mol. Phys.}\ }\textbf {\bibinfo {volume}
  {17}},\ \bibinfo {pages} {4223} (\bibinfo {year} {1984})}\BibitemShut
  {NoStop}%
\bibitem [{\citenamefont {Massignan}(2012)}]{Massignan:2012_narrow}%
  \BibitemOpen
  \bibfield  {author} {\bibinfo {author} {\bibfnamefont {P.}~\bibnamefont
  {Massignan}},\ }\bibfield  {title} {\bibinfo {title} {{Polarons and dressed
  molecules near narrow Feshbach resonances}},\ }\href
  {https://doi.org/10.1209/0295-5075/98/10012} {\bibfield  {journal} {\bibinfo
  {journal} {EPL}\ }\textbf {\bibinfo {volume} {98}},\ \bibinfo {pages} {10012}
  (\bibinfo {year} {2012})}\BibitemShut {NoStop}%
\bibitem [{\citenamefont {J{\'o}{\'z}wiak}\ \emph {et~al.}(2026)\citenamefont
  {J{\'o}{\'z}wiak}, \citenamefont {Yang}, \citenamefont {Dizer}, \citenamefont
  {Christianen},\ and\ \citenamefont {Karman}}]{jozwiak:2026}%
  \BibitemOpen
  \bibfield  {author} {\bibinfo {author} {\bibfnamefont {H.~J.}\ \bibnamefont
  {J{\'o}{\'z}wiak}}, \bibinfo {author} {\bibfnamefont {H.}~\bibnamefont
  {Yang}}, \bibinfo {author} {\bibfnamefont {E.}~\bibnamefont {Dizer}},
  \bibinfo {author} {\bibfnamefont {A.}~\bibnamefont {Christianen}},\ and\
  \bibinfo {author} {\bibfnamefont {T.}~\bibnamefont {Karman}},\ }\bibfield
  {title} {\bibinfo {title} {Tunable state-dependent interactions in
  collisionally stable mixtures of polar molecules},\ }\href@noop {} {\
  (\bibinfo {year} {2026})},\ \Eprint {https://arxiv.org/abs/2607.25777}
  {arXiv:2607.25777 [cond-mat.quant-gas]} \BibitemShut {NoStop}%
\bibitem [{\citenamefont {Yang}\ \emph {et~al.}(2026)\citenamefont {Yang},
  \citenamefont {Dizer}, \citenamefont {Schmidt}, \citenamefont {Christianen},\
  and\ \citenamefont {Karman}}]{Yang:2026}%
  \BibitemOpen
  \bibfield  {author} {\bibinfo {author} {\bibfnamefont {H.}~\bibnamefont
  {Yang}}, \bibinfo {author} {\bibfnamefont {E.}~\bibnamefont {Dizer}},
  \bibinfo {author} {\bibfnamefont {R.}~\bibnamefont {Schmidt}}, \bibinfo
  {author} {\bibfnamefont {A.}~\bibnamefont {Christianen}},\ and\ \bibinfo
  {author} {\bibfnamefont {T.}~\bibnamefont {Karman}},\ }\bibfield  {title}
  {\bibinfo {title} {Polarons in microwave-shielded vibrational mixtures of
  ultracold molecules}} (\bibinfo {year} {2026}),\ \bibinfo {note} {manuscript
  in preparation}\BibitemShut {NoStop}%
\bibitem [{\citenamefont {Cetina}\ \emph {et~al.}(2016)\citenamefont {Cetina},
  \citenamefont {Jag}, \citenamefont {Lous}, \citenamefont {Fritsche},
  \citenamefont {Walraven}, \citenamefont {Grimm}, \citenamefont {Levinsen},
  \citenamefont {Parish}, \citenamefont {Schmidt}, \citenamefont {Knap},\ and\
  \citenamefont {Demler}}]{Cetina:2016}%
  \BibitemOpen
  \bibfield  {author} {\bibinfo {author} {\bibfnamefont {M.}~\bibnamefont
  {Cetina}}, \bibinfo {author} {\bibfnamefont {M.}~\bibnamefont {Jag}},
  \bibinfo {author} {\bibfnamefont {R.~S.}\ \bibnamefont {Lous}}, \bibinfo
  {author} {\bibfnamefont {I.}~\bibnamefont {Fritsche}}, \bibinfo {author}
  {\bibfnamefont {J.~T.~M.}\ \bibnamefont {Walraven}}, \bibinfo {author}
  {\bibfnamefont {R.}~\bibnamefont {Grimm}}, \bibinfo {author} {\bibfnamefont
  {J.}~\bibnamefont {Levinsen}}, \bibinfo {author} {\bibfnamefont {M.~M.}\
  \bibnamefont {Parish}}, \bibinfo {author} {\bibfnamefont {R.}~\bibnamefont
  {Schmidt}}, \bibinfo {author} {\bibfnamefont {M.}~\bibnamefont {Knap}},\ and\
  \bibinfo {author} {\bibfnamefont {E.}~\bibnamefont {Demler}},\ }\bibfield
  {title} {\bibinfo {title} {{Ultrafast many-body interferometry of impurities
  coupled to a Fermi sea}},\ }\href {https://doi.org/10.1126/science.aaf5134}
  {\bibfield  {journal} {\bibinfo  {journal} {Science}\ }\textbf {\bibinfo
  {volume} {354}},\ \bibinfo {pages} {96} (\bibinfo {year} {2016})}\BibitemShut
  {NoStop}%
\bibitem [{\citenamefont {Goussev}\ \emph {et~al.}(2012)\citenamefont
  {Goussev}, \citenamefont {Jalabert}, \citenamefont {Pastawski},\ and\
  \citenamefont {Wisniacki}}]{Goussev:2012}%
  \BibitemOpen
  \bibfield  {author} {\bibinfo {author} {\bibfnamefont {A.}~\bibnamefont
  {Goussev}}, \bibinfo {author} {\bibfnamefont {R.~A.}\ \bibnamefont
  {Jalabert}}, \bibinfo {author} {\bibfnamefont {H.~M.}\ \bibnamefont
  {Pastawski}},\ and\ \bibinfo {author} {\bibfnamefont {D.}~\bibnamefont
  {Wisniacki}},\ }\bibfield  {title} {\bibinfo {title} {{Loschmidt echo}},\
  }\href {https://doi.org/10.4249/scholarpedia.11687} {\bibfield  {journal}
  {\bibinfo  {journal} {Scholarpedia}\ }\textbf {\bibinfo {volume} {7}},\
  \bibinfo {pages} {11687} (\bibinfo {year} {2012})}\BibitemShut {NoStop}%
\bibitem [{\citenamefont {Levinsen}\ \emph {et~al.}(2012)\citenamefont
  {Levinsen}, \citenamefont {Massignan}, \citenamefont {Chevy},\ and\
  \citenamefont {Lobo}}]{PhysRevLett.109.075302}%
  \BibitemOpen
  \bibfield  {author} {\bibinfo {author} {\bibfnamefont {J.}~\bibnamefont
  {Levinsen}}, \bibinfo {author} {\bibfnamefont {P.}~\bibnamefont {Massignan}},
  \bibinfo {author} {\bibfnamefont {F.}~\bibnamefont {Chevy}},\ and\ \bibinfo
  {author} {\bibfnamefont {C.}~\bibnamefont {Lobo}},\ }\bibfield  {title}
  {\bibinfo {title} {{$p$-Wave Polaron}},\ }\href
  {https://doi.org/10.1103/PhysRevLett.109.075302} {\bibfield  {journal}
  {\bibinfo  {journal} {Phys. Rev. Lett.}\ }\textbf {\bibinfo {volume} {109}},\
  \bibinfo {pages} {075302} (\bibinfo {year} {2012})}\BibitemShut {NoStop}%
\bibitem [{Note1()}]{Note1}%
  \BibitemOpen
  \bibinfo {note} {The odd-wave onset sits at $1/(k_Fa) = (a_d/a)_c/(k_Fa_d)$,
  so it approaches resonance as $k_Fa_d$ grows.}\BibitemShut {Stop}%
\bibitem [{Note2()}]{Note2}%
  \BibitemOpen
  \bibinfo {note} {Our convention $a_d = m_r d^2/\hbar ^2$ gives $a_d = m\mu
  _0\mu ^2/(8\pi \hbar ^2)$ with $d^2 = \mu _0\mu ^2/(4\pi )$. This differs by
  a factor of $3/2$ from the bosonic convention $a_{\protect \rm dd} = m\mu
  _0\mu ^2/(12\pi \hbar ^2)$~\cite {Chomaz:2023}, under which dysprosium has
  $a_{\protect \rm dd}\approx 130\protect \,a_0$.}\BibitemShut {Stop}%
\bibitem [{\citenamefont {Soave}\ \emph {et~al.}(2022)\citenamefont {Soave},
  \citenamefont {Corre}, \citenamefont {Ravensbergen}, \citenamefont {Han},
  \citenamefont {Kreyer}, \citenamefont {Kirilov},\ and\ \citenamefont
  {Grimm}}]{soave:2022}%
  \BibitemOpen
  \bibfield  {author} {\bibinfo {author} {\bibfnamefont {E.}~\bibnamefont
  {Soave}}, \bibinfo {author} {\bibfnamefont {V.}~\bibnamefont {Corre}},
  \bibinfo {author} {\bibfnamefont {C.}~\bibnamefont {Ravensbergen}}, \bibinfo
  {author} {\bibfnamefont {J.~H.}\ \bibnamefont {Han}}, \bibinfo {author}
  {\bibfnamefont {M.}~\bibnamefont {Kreyer}}, \bibinfo {author} {\bibfnamefont
  {E.}~\bibnamefont {Kirilov}},\ and\ \bibinfo {author} {\bibfnamefont
  {R.}~\bibnamefont {Grimm}},\ }\bibfield  {title} {\bibinfo {title}
  {{Low-field Feshbach resonances and three-body losses in a fermionic quantum
  gas of $^{161}${D}y}},\ }\href
  {https://doi.org/https://doi.org/10.15407/ujpe67.5.334} {\bibfield  {journal}
  {\bibinfo  {journal} {Ukr. J. Phys.}\ }\textbf {\bibinfo {volume} {67}},\
  \bibinfo {pages} {334} (\bibinfo {year} {2022})}\BibitemShut {NoStop}%
\bibitem [{\citenamefont {Wang}\ \emph {et~al.}(2024)\citenamefont {Wang},
  \citenamefont {Biswas}, \citenamefont {Eppelt}, \citenamefont {Deng},
  \citenamefont {Luo},\ and\ \citenamefont {Bohn}}]{Wang:2024}%
  \BibitemOpen
  \bibfield  {author} {\bibinfo {author} {\bibfnamefont {R.~R.~W.}\
  \bibnamefont {Wang}}, \bibinfo {author} {\bibfnamefont {S.}~\bibnamefont
  {Biswas}}, \bibinfo {author} {\bibfnamefont {S.}~\bibnamefont {Eppelt}},
  \bibinfo {author} {\bibfnamefont {F.}~\bibnamefont {Deng}}, \bibinfo {author}
  {\bibfnamefont {X.-Y.}\ \bibnamefont {Luo}},\ and\ \bibinfo {author}
  {\bibfnamefont {J.~L.}\ \bibnamefont {Bohn}},\ }\bibfield  {title} {\bibinfo
  {title} {{Simulations of evaporation to deep Fermi degeneracy in
  microwave-shielded molecules}},\ }\href
  {https://doi.org/10.1103/PhysRevA.110.043309} {\bibfield  {journal} {\bibinfo
   {journal} {Phys. Rev. A}\ }\textbf {\bibinfo {volume} {110}},\ \bibinfo
  {pages} {043309} (\bibinfo {year} {2024})}\BibitemShut {NoStop}%
\bibitem [{\citenamefont {Nozi\`eres}\ and\ \citenamefont
  {De~Dominicis}(1969)}]{Nozieres:1969}%
  \BibitemOpen
  \bibfield  {author} {\bibinfo {author} {\bibfnamefont {P.}~\bibnamefont
  {Nozi\`eres}}\ and\ \bibinfo {author} {\bibfnamefont {C.~T.}\ \bibnamefont
  {De~Dominicis}},\ }\bibfield  {title} {\bibinfo {title} {{Singularities in
  the X-Ray Absorption and Emission of Metals. III. One-Body Theory Exact
  Solution}},\ }\href {https://doi.org/10.1103/PhysRev.178.1097} {\bibfield
  {journal} {\bibinfo  {journal} {Phys. Rev.}\ }\textbf {\bibinfo {volume}
  {178}},\ \bibinfo {pages} {1097} (\bibinfo {year} {1969})}\BibitemShut
  {NoStop}%
\bibitem [{\citenamefont {Combescot}\ and\ \citenamefont
  {Nozi\`eres}(1971)}]{Combescot:1971}%
  \BibitemOpen
  \bibfield  {author} {\bibinfo {author} {\bibfnamefont {M.}~\bibnamefont
  {Combescot}}\ and\ \bibinfo {author} {\bibfnamefont {P.}~\bibnamefont
  {Nozi\`eres}},\ }\bibfield  {title} {\bibinfo {title} {{Infrared catastrophy
  and excitons in the X-ray spectra of metals}},\ }\href
  {https://doi.org/10.1051/jphys:019710032011-12091300} {\bibfield  {journal}
  {\bibinfo  {journal} {J. Phys. France}\ }\textbf {\bibinfo {volume} {32}},\
  \bibinfo {pages} {913} (\bibinfo {year} {1971})}\BibitemShut {NoStop}%
\bibitem [{\citenamefont {Chen}\ \emph {et~al.}(2025)\citenamefont {Chen},
  \citenamefont {Dizer}, \citenamefont {Rodr\'{\i}guez},\ and\ \citenamefont
  {Schmidt}}]{Chen:2025}%
  \BibitemOpen
  \bibfield  {author} {\bibinfo {author} {\bibfnamefont {X.}~\bibnamefont
  {Chen}}, \bibinfo {author} {\bibfnamefont {E.}~\bibnamefont {Dizer}},
  \bibinfo {author} {\bibfnamefont {E.~R.}\ \bibnamefont {Rodr\'{\i}guez}},\
  and\ \bibinfo {author} {\bibfnamefont {R.}~\bibnamefont {Schmidt}},\
  }\bibfield  {title} {\bibinfo {title} {{Mass-Gap Description of Heavy
  Impurities in Fermi Gases}},\ }\href {https://doi.org/10.1103/h2f7-dhjh}
  {\bibfield  {journal} {\bibinfo  {journal} {Phys. Rev. Lett.}\ }\textbf
  {\bibinfo {volume} {135}},\ \bibinfo {pages} {193401} (\bibinfo {year}
  {2025})}\BibitemShut {NoStop}%
\bibitem [{\citenamefont {Peng}\ \emph {et~al.}(2021)\citenamefont {Peng},
  \citenamefont {Liu}, \citenamefont {Zhang},\ and\ \citenamefont
  {Cui}}]{PhysRevA.103.063312}%
  \BibitemOpen
  \bibfield  {author} {\bibinfo {author} {\bibfnamefont {C.}~\bibnamefont
  {Peng}}, \bibinfo {author} {\bibfnamefont {R.}~\bibnamefont {Liu}}, \bibinfo
  {author} {\bibfnamefont {W.}~\bibnamefont {Zhang}},\ and\ \bibinfo {author}
  {\bibfnamefont {X.}~\bibnamefont {Cui}},\ }\bibfield  {title} {\bibinfo
  {title} {{Nature of the polaron-molecule transition in Fermi polarons}},\
  }\href {https://doi.org/10.1103/PhysRevA.103.063312} {\bibfield  {journal}
  {\bibinfo  {journal} {Phys. Rev. A}\ }\textbf {\bibinfo {volume} {103}},\
  \bibinfo {pages} {063312} (\bibinfo {year} {2021})}\BibitemShut {NoStop}%
\bibitem [{\citenamefont {Read}\ and\ \citenamefont {Green}(2000)}]{Read:2000}%
  \BibitemOpen
  \bibfield  {author} {\bibinfo {author} {\bibfnamefont {N.}~\bibnamefont
  {Read}}\ and\ \bibinfo {author} {\bibfnamefont {D.}~\bibnamefont {Green}},\
  }\bibfield  {title} {\bibinfo {title} {{Paired states of fermions in two
  dimensions with breaking of parity and time-reversal symmetries and the
  fractional quantum Hall effect}},\ }\href
  {https://doi.org/10.1103/PhysRevB.61.10267} {\bibfield  {journal} {\bibinfo
  {journal} {Phys. Rev. B}\ }\textbf {\bibinfo {volume} {61}},\ \bibinfo
  {pages} {10267} (\bibinfo {year} {2000})}\BibitemShut {NoStop}%
\bibitem [{\citenamefont {Gurarie}\ \emph {et~al.}(2005)\citenamefont
  {Gurarie}, \citenamefont {Radzihovsky},\ and\ \citenamefont
  {Andreev}}]{Gurarie:2005}%
  \BibitemOpen
  \bibfield  {author} {\bibinfo {author} {\bibfnamefont {V.}~\bibnamefont
  {Gurarie}}, \bibinfo {author} {\bibfnamefont {L.}~\bibnamefont
  {Radzihovsky}},\ and\ \bibinfo {author} {\bibfnamefont {A.~V.}\ \bibnamefont
  {Andreev}},\ }\bibfield  {title} {\bibinfo {title} {{Quantum Phase
  Transitions across a $p$-Wave Feshbach Resonance}},\ }\href
  {https://doi.org/10.1103/PhysRevLett.94.230403} {\bibfield  {journal}
  {\bibinfo  {journal} {Phys. Rev. Lett.}\ }\textbf {\bibinfo {volume} {94}},\
  \bibinfo {pages} {230403} (\bibinfo {year} {2005})}\BibitemShut {NoStop}%
\bibitem [{\citenamefont {Christianen}(2023)}]{Christianen:2023qig}%
  \BibitemOpen
  \bibfield  {author} {\bibinfo {author} {\bibfnamefont {A.}~\bibnamefont
  {Christianen}},\ }\emph {\bibinfo {title} {{Chemistry in a quantum
  medium}}},\ \href {https://mediatum.ub.tum.de/1720019} {Ph.D. thesis},\
  \bibinfo  {school} {TU Munich} (\bibinfo {year} {2023})\BibitemShut {NoStop}%
\bibitem [{\citenamefont {Haydock}\ \emph {et~al.}(1972)\citenamefont
  {Haydock}, \citenamefont {Heine},\ and\ \citenamefont {Kelly}}]{Haydock1972}%
  \BibitemOpen
  \bibfield  {author} {\bibinfo {author} {\bibfnamefont {R.}~\bibnamefont
  {Haydock}}, \bibinfo {author} {\bibfnamefont {V.}~\bibnamefont {Heine}},\
  and\ \bibinfo {author} {\bibfnamefont {M.~J.}\ \bibnamefont {Kelly}},\
  }\bibfield  {title} {\bibinfo {title} {{Electronic structure based on the
  local atomic environment for tight-binding bands}},\ }\href
  {https://doi.org/10.1088/0022-3719/5/20/004} {\bibfield  {journal} {\bibinfo
  {journal} {J. Phys. C: Solid State Phys.}\ }\textbf {\bibinfo {volume} {5}},\
  \bibinfo {pages} {2845} (\bibinfo {year} {1972})}\BibitemShut {NoStop}%
\bibitem [{\citenamefont {Tiene}\ \emph {et~al.}(2022)\citenamefont {Tiene},
  \citenamefont {Levinsen}, \citenamefont {Keeling}, \citenamefont {Parish},\
  and\ \citenamefont {Marchetti}}]{PhysRevB.105.125404}%
  \BibitemOpen
  \bibfield  {author} {\bibinfo {author} {\bibfnamefont {A.}~\bibnamefont
  {Tiene}}, \bibinfo {author} {\bibfnamefont {J.}~\bibnamefont {Levinsen}},
  \bibinfo {author} {\bibfnamefont {J.}~\bibnamefont {Keeling}}, \bibinfo
  {author} {\bibfnamefont {M.~M.}\ \bibnamefont {Parish}},\ and\ \bibinfo
  {author} {\bibfnamefont {F.~M.}\ \bibnamefont {Marchetti}},\ }\bibfield
  {title} {\bibinfo {title} {{Effect of fermion indistinguishability on optical
  absorption of doped two-dimensional semiconductors}},\ }\href
  {https://doi.org/10.1103/PhysRevB.105.125404} {\bibfield  {journal} {\bibinfo
   {journal} {Phys. Rev. B}\ }\textbf {\bibinfo {volume} {105}},\ \bibinfo
  {pages} {125404} (\bibinfo {year} {2022})}\BibitemShut {NoStop}%
\end{thebibliography}
\end{document}